\documentclass[10pt,amsmath,eqsecnum,prd]{revtex4}
\newcommand{\eq}[1]{Eq.~{(\ref{#1})}}
\newcommand{\fig}[1]{Fig.~{\ref{#1}}}
\newcommand{\bea}{\begin{eqnarray}}
\newcommand{\beann}{\begin{eqnarray*}}
\newcommand{\eea}{\end{eqnarray}}
\newcommand{\eeann}{\end{eqnarray*}}

\usepackage{graphicx}
\usepackage{latexsym}
\usepackage{amssymb}
\usepackage{amsbsy}
\DeclareMathAlphabet{\mathsfsl}{OT1}{cmss}{m}{sl}
\newcommand{\bs}{\boldsymbol}
\renewcommand{\vec}[1]{\bs{#1}}

\newcommand{\cvec}[1]{\overstar{\vec{#1}}}  % Complex vector
\newcommand{\vect}[1]{{}^3\vec{#1}}
\newcommand{\mv}[1]{\mathbf{#1}}
\newcommand{\pv}[1]{\underline{\vec{#1}}}
\renewcommand{\tensor}[1]{\mathsfsl{#1}}
\newcommand{\op}[1]{\underline{\tensor{#1}}}
\newcommand{\inv}[1]{\op{#1}^{-1}}
\newcommand{\adj}[1]{\overline{\tensor{#1}}}

\newcommand{\Cl}{\mathcal{C}\ell}
\newcommand{\Ccl}{\mathbb{C}\ell}

\newcommand{\grade}[1]{\left\langle #1 \right\rangle}

\newcommand{\SL}{\text{SL}}
\newcommand{\SU}{\text{SU}}
\newcommand{\SO}{\text{SO}}
\newcommand{\spin}{\text{Spin}}

\begin{document}
\title{The Construction of Spinors in Geometric Algebra}
\author{Matthew R. Francis} \email{mfrancis@physics.rutgers.edu}
\author{Arthur Kosowsky} \email{kosowsky@physics.rutgers.edu} \affiliation{Dept. of Physics and Astronomy, Rutgers University\\136 Frelinghuysen Road, Piscataway, NJ 08854}
\date{\today}

\begin{abstract}
The relationship between spinors and Clifford (or geometric) algebra has long been studied, but little consistency may be found between the various approaches.  However, when spinors are defined to be elements of the even subalgebra of some real geometric algebra, the gap between algebraic, geometric, and physical methods is closed.  Spinors are developed in any number of dimensions from a discussion of spin groups, followed by the specific cases of $\text{U}(1)$, $\SU(2)$, and $\text{SL}(2,\mathbb{C})$ spinors.  The physical observables in Schr\"{o}dinger-Pauli theory and Dirac theory are found, and the relationship between Dirac, Lorentz, Weyl, and Majorana spinors is made explicit.  The use of a real geometric algebra, as opposed to one defined over the complex numbers, provides a simpler construction and advantages of conceptual and theoretical clarity not available in other approaches.
\end{abstract}
\maketitle

\section{Introduction}

Spinors are used in a wide range of fields, from the quantum physics of fermions and general relativity, to fairly abstract areas of algebra and geometry.  Independent of the particular application, the defining characteristic of spinors is their behavior under rotations:  for a given angle $\theta$ that a vector or tensorial object rotates, a spinor rotates by $\theta/2$, and hence takes two full rotations to return to its original configuration.  The \emph{spin groups}, which are universal coverings of the rotation groups, govern this behavior, and are frequently defined in the language of geometric (Clifford) algebras \cite{porteous,lounesto}.

In this paper, we follow the geometric algebra approach to its logical conclusion, and define spinors as arbitrary elements of the \emph{even subalgebra} of a real geometric algebra; since the spin group is made up of normalized even multivectors, the action of a rotation maps spinor space onto itself.  The identification of spinors with real even multivectors in geometric algebra was first made by David Hestenes \cite{realspinors,hest86,CA2GC}; we extend his approach using group theory and insights provided by algebraic spinor methods (see \emph{e.g.} Ref.~[\onlinecite{crumeyrolle}]).  

Many modern mathematical treatments (Refs.~[\onlinecite{spin_geom,gockeler}], for example) begin by defining a \emph{complex} geometric algebra, in which the representation of the spin group lives.  Spinors are then written as members of left minimal ideals of the Clifford algebra.  Although this method is closely related to our treatment, the geometrical interpretation is muddied by the presence of the imaginary unit $i$.  In addition, spinors in the left-ideal approach lie in the full geometric algebra, rather than its even subset, and the specific left minimal ideal is dependent on the full algebra.  By contrast, if spinors are assigned to the same algebra that defines the spin group, they may be embedded in algebras of higher dimension.  Left ideals do continue to play a role, however, and are used in relating Dirac spinors to two-component (Lorentz) spinors in spacetime.

The advantage of defining spinors over the field of real numbers, however, lies in the fact that every multivector in a real geometric algebra has a geometrical interpretation (however complicated).  In fact, the introduction of imaginary units in spinor theory arose out of matrix representations of spin groups, but when the same group is represented in a real geometric algebra, the introduction of complex numbers is superfluous.  (The relationship between matrix and geometric algebra approaches to group theory is obtained by representing a geometric algebra as a matrix algebra; see Appendix~\ref{matrep}.)  The complex (Hermitian) structure of spinors is found by specifying a ``spin-axis'' in the space under consideration, so that it depends not only on the dimensionality of the group, but also on an orientation for the space.  Although we will not discuss topics in topology in this paper, the existence of such a complex structure seems to be intimately related to the existence of global spin structure in a manifold (see Refs.~[\onlinecite{penrose_rindler1,benn}]).  Since defining spinors in a real algebra is extensible to any dimension, all the results of standard spinor theory can be carried over.

Section~\ref{spingroups} introduces spin groups for any dimension and signature in the language of geometric algebra, defining rotations in terms of multivector objects.  The next sections deal with specific spin groups, and their associated spinors:  Section~\ref{u1} treats two-dimensional Euclidean and anti-Euclidean spinors, Section~\ref{su2} is concerned with Pauli spinors in three-dimensions, while spacetime spinors are the subject of Sections~\ref{sl2c} and \ref{lorentzspin}.  In each case, we will show that the complex structure deemed necessary to treat spinors arises naturally from the real geometric algebra; the connection to complex geometric algebras is made in Section~\ref{ideals} and Appendix~\ref{matrep}.  The physical properties of Pauli, Dirac, and Lorentz spinors are developed along the way, in a manner that elucidates their geometry.

\section{Geometric Algebra and Spin Groups}
\label{spingroups}

Spinors were originally discovered in the context of rotation group representations in three dimensions \cite{cartan_spin}, and quickly found applications in quantum theory and general relativity.  From the beginning, the relationship between geometric algebras and spinors has been known \cite{cartan_spin,chevalley,riesz}, although often via matrix representations.  This section first gives a brief synopsis of the properties of geometric algebra relevant to spinors, then defines spin groups in this language.

A (real) geometric algebra $\Cl_{p,q}$ is characterized by its total (vector) dimension $n=p+q$ and its signature $s=p-q$, where $p$ is the number of basis vectors with positive norm and $q$ enumerates the basis vectors with negative norm.  In a suitable orthonormal basis, the norm of a vector is written as
\bea
\vec{a}^2 = \sum^p_{k=1} a^2_k - \sum^{p+q}_{l=p+1} a^2_l ,
\label{quadratic}
\eea
which is scalar-valued.  A generic multivector is written as the sum of its grades
\bea
\mv{A} = \grade{\mv{A}}_0 + \grade{\mv{A}}_1 + \grade{\mv{A}}_2 + \ldots + \grade{\mv{A}}_{p+q} = \sum^{p+q}_{k=0} \grade{\mv{A}}_k
\eea
where $\grade{\mv{A}}_0$ is the scalar part, $\grade{\mv{A}}_1$ is the vector part, $\grade{\mv{A}}_2$ is the bivector part, and so forth.  As usual, a bivector is the outer product of two vectors, or the linear combination of such objects:
\bea
\grade{\mv{B}}_2 = \sum \vec{a}_k \wedge \vec{b}_k .
\eea
The total number of linearly-independent multivectors in $\Cl_{p,q}$ is $2^{p+q}$.  (Refer to Doran and Lasenby\cite{GA03} for a practical introduction to geometric algebras, and Lounesto\cite{lounesto} for a thorough mathematical treatment.)

The \emph{main involution} changes the sign of all vectors in the algebra (without changing the order of multiplication):
\bea
\widehat{\mv{AB}} = \hat{\mv{A}}\hat{\mv{B}} \qquad \to \quad \widehat{\grade{\vec{a}}}_1 = - \grade{\vec{a}}_1 ,
\eea
so that
\bea
\hat{\mv{A}} = \grade{\mv{A}}_0 - \grade{\mv{A}}_1 + \grade{\mv{A}}_2 - \ldots + (-1)^{p+q} \grade{\mv{A}}_{p+q} = \sum^{p+q}_{k=0} (-1)^k \grade{\mv{A}}_k
\eea
for a general multivector.  This operation splits the algebra into subspaces, called even and odd:
\bea
\grade{\mv{A}}_\pm = \frac{1}{2} \left( \mv{A} \pm \hat{\mv{A}} \right) \ \in \Cl^\pm_{p,q}.
\eea
The even multivectors form an algebra under the geometric product, which we call the even subalgebra $\Cl^+_{p,q}$, which is independent of signature $s = p-q$:  $\Cl^+_{p,q} \simeq \Cl^+_{q,p}$.  The even subalgebra is isomorphic to a geometric algebra of smaller dimension \cite{benn,lounesto}:
\beann
\Cl^+_{p,q} \simeq \Cl_{q,p-1} \simeq \Cl_{p,q-1};
\eeann
in general, note that $\Cl_{p,q} \simeq \Cl_{q+1,p-1}$.  The odd multivectors do not form an algebra, but the following rules apply:
\bea
\grade{\mv{A}}_+ \grade{\mv{B}}_+ &\in& \Cl^+_{p,q} \nonumber \\
\grade{\mv{A}}_- \grade{\mv{B}}_- &\in& \Cl^+_{p,q} \nonumber \\
\grade{\mv{A}}_+ \grade{\mv{B}}_- &\in& \Cl^-_{p,q}.
\eea

The \emph{reverse} preserves the sign of vectors, but reverses the order of multiplication:
\bea
\widetilde{\mv{AB}} = \tilde{\mv{B}}\tilde{\mv{A}} \qquad \to \quad \widetilde{\grade{\vec{a}}}_1 = \grade{\vec{a}}_1 ,
\eea
so that
\bea
\tilde{\mv{A}} = \grade{\mv{A}}_0 + \grade{\mv{A}}_1 - \grade{\mv{A}}_2 - \ldots + (-1)^{(p+q)(p+q-1)/2}\grade{\mv{A}}_{p+q} = \sum^{p+q}_{k=0} (-1)^{k(k-1)/2} \grade{\mv{A}}_k
\eea
for a generic multivector.  The reverse is identical to Hermitian conjugation in a Euclidean geometric algebra $\Cl_n = \Cl_{n,0}$. (In fact, Hestenes and Sobczyk\cite{CA2GC}, who work mostly in Euclidean spaces, use the notation $\mv{A}^\dagger$ for reversion.)  We can also combine the reverse and the main involution to make the \emph{Clifford conjugate} \cite{lounesto}:
\bea
\bar{\mv{A}} \equiv \hat{\tilde{\mv{A}}} = \sum^{p+q}_{k=0} (-1)^{k(k+1)/2} \grade{\mv{A}}_k ,
\eea
which both reverses the order of multiplication and changes the sign of vectors.

Consider an operation $\op{\Gamma}$ that maps a vector onto another vector, and preserves the norm:
\bea
\op{\Gamma}(\vec{a}) = \vec{a}' \qquad \to \qquad \vec{a}'{}^2 = \op{\Gamma}(\vec{a}) \op{\Gamma}(\vec{a}) = \adj{\Gamma}\left(\op{\Gamma}(\vec{a})\right)\vec{a} = \vec{a}^2
\eea
where $\adj{\Gamma}$ is the adjoint of $\op{\Gamma}$.  Since $\adj{\Gamma} = \inv{\Gamma}$, this operator is orthogonal and thus is in the group $\text{O}(p,q)$.  Such an operator can be represented in $\Cl_{p,q}$ as
\bea
\op{\Gamma}(\vec{a}) = \pm \mv{U}\vec{a}\mv{U}^{-1} ,
\label{pin}
\eea
where $\mv{U}$ is an as-yet unspecified invertible multivector.  To ensure that $\op{\Gamma}(\vec{a})$ is a vector, we have the condition
\bea
\widetilde{\op{\Gamma}(\vec{a})} = \op{\Gamma}(\vec{a}) \qquad \to \qquad \mv{U}\tilde{\mv{U}} = \pm 1
\eea
The invertible multivectors $\tilde{\mv{U}} = \mv{U}^{-1}$ form the group $\text{Pin}(p,q)$, which is a double covering of $\text{O}(p,q)$ \cite{porteous}.  This group can also be extended to act on arbitrary multivectors, in which case we have the general identity
\bea
\op{\Gamma}(\mv{A})\op{\Gamma}(\mv{B}) = \op{\Gamma}(\mv{AB}) \quad \text{where} \quad \op{\Gamma}(\mv{A}) = \pm \mv{U} \mv{A} \tilde{\mv{U}}.
\eea
The Pin groups contain rotations and reflections, along with other operations.

Since even multivectors form an algebra, the Pin group contains a subgroup comprised only of invertible even multivectors, which is $\spin(p,q)$:
\bea
\op{R}(\mv{A}) = \grade{\mv{U}}_+ \mv{A} \widetilde{\grade{\mv{U}}}_+ \quad \text{where} \quad \grade{\mv{U}}_+ \widetilde{\grade{\mv{U}}}_+ = \pm 1 .
\eea
The proper subgroup $\spin_+(p,q)$ contains the elements that normalize to unity.  $\spin(p,q)$ and $\spin_+(p,q)$ are double coverings of the groups $\SO(p,q)$ and $\SO_+(p,q)$, respectively, which are the rotation groups.  Since $\Cl^+_{p,q} \simeq \Cl^+_{q,p}$, we have the group isomorphism $\spin(p,q)\simeq\spin(q,p)$ (a fact we will exploit in Appendix~\ref{stsig}), which means spin groups for Euclidean and anti-Euclidean spaces of the same dimension are isomorphic; in addition, $\spin(n)=\spin(n,0)=\spin_+(n)$.  The (positive or negative) exponential of a bivector is always in the proper spin group:
\bea
\pm \text{e}^{\grade{\mv{B}}_2} = \pm \left( 1 + \grade{\mv{B}}_2 + \frac{1}{2} \grade{\mv{B}}_2{}^2 + \frac{1}{3!} \grade{\mv{B}}_2{}^3 + \ldots \right) = \pm \sum^\infty_{r=0} \frac{1}{r!} \grade{\mv{B}}_2{}^r \in \spin_+(p,q),\nonumber\\
\label{exponent}
\eea
where $\grade{\mv{B}}_2{}^2 = \grade{\mv{B}}_2\grade{\mv{B}}_2$ and so forth.  Both signs are necessary:  when $\grade{\mv{B}}_2{}^2 = 0$, for example, it is not possible to find a bivector $\grade{\mv{B}'}_2$ such that $-\exp(\grade{\mv{B}}_2) = \exp(\grade{\mv{B}'}_2)$ \cite{lounesto}.  For the main examples we discuss, it is always possible to write members of the spin group in the form of \eq{exponent}; however, some groups do not allow such a representation.

A \emph{spinor} is an object that transforms under one-sided multiplication by an element of a spin group.  In other words, if we let $\Sigma_{p,q}$ be the (left-invariant) spinor space associated with $\Cl_{p,q}$, the following relation holds:
\bea
\psi' = \mv{U} \psi \qquad \forall \mv{U} \in \spin(p,q) \quad \text{and} \quad \psi,\psi' \in \Sigma_{p,q} .
\label{spinordef}
\eea
The most obvious choice for $\Sigma_{p,q}$ is the set of even multivectors $\Cl^+_{p,q}$, and in fact we will make this identification; the relationship between the even multivector approach and the algebraic approach of left minimal ideals is made in Section~\ref{ideals} and \ref{discussion}.  Note that the product \[ \tilde{\psi}'\psi' = \tilde{\psi}\tilde{\mv{U}}\mv{U}\psi = \tilde{\psi}\psi \] is \emph{spin-invariant}.

The next sections are devoted to specific examples of spin groups, and their associated spinors.

\section{Two Dimensions: Complex Numbers and $\text{U}(1)$ Spinors}
\label{u1}

\subsection{The Euclidean Algebra $\Cl_2$}
\label{cl2}

The simplest nontrivial rotation group occurs in the two-dimensional Euclidean geometric algebra $\Cl_2$, to which we assign the multivector basis
\bea
1,\ \vec{e}_1,\ \vec{e}_2,\ \mv{e}_{12} \equiv \vec{e}_1\vec{e}_2
\eea
so that the product of the basis vectors may be written as
\bea
\vec{e}_i \vec{e}_j = \grade{\vec{e}_i \vec{e}_j}_0 + \grade{\vec{e}_i \vec{e}_j}_2 = \delta_{ij} + \epsilon_{ij} \mv{e}_{12}.
\eea
The bivector (or pseudoscalar) element $\mv{e}_{12}$ anticommutes with every vector:  $\vec{a} \mv{e}_{12} = - \mv{e}_{12} \vec{a}$.

The even subalgebra $\Cl^+_2 \simeq \Cl_{0,1}$ is isomorphic to the algebra of complex numbers, with the reverse as the conjugation operation:
\bea
\grade{\mv{A}}_+ = \grade{\mv{A}}_0 + \grade{\mv{A}}_2 \quad \text{where} \quad \grade{\mv{A}}_+ \grade{\mv{B}}_+ = \grade{\mv{B}}_+ \grade{\mv{A}}_+ \nonumber \\
\to \quad \grade{\mv{A}}_+ \widetilde{\grade{\mv{A}}}_+ = \grade{\mv{A}}_0{}^2 + \grade{\mv{A}}_2{}^2
\eea
and $\mv{e}_{12}{}^2 = -1$.  The spin group $\spin(2)=\spin_+(2)$ is therefore comprised of the unitary ``complex numbers'', and as such is equivalent to the one-dimensional unitary group $\text{U}(1)$.  Since all bivectors in $\Cl_2$ are proportional to $\mv{e}_{12}$, all unitary even multivectors can be written as
\beann
\mv{U} = \text{e}^{\theta \mv{e}_{12}/2} = \cos (\theta/2) + \mv{e}_{12} \sin (\theta/2) \quad \text{and} \quad \tilde{\mv{U}} = \text{e}^{-\theta \mv{e}_{12}/2} = \cos (\theta/2) - \mv{e}_{12} \sin (\theta/2),
\eeann
where $\theta$ is a scalar angle.  Due to the commutative nature of $\Cl^+_2$, $\spin(2)$ is an Abelian group.

For a vector in $\Cl_2$, a rotation is written as
\bea
\op{R}(\vec{a}) = \text{e}^{\theta \mv{e}_{12}/2} \vec{a} \text{e}^{-\theta \mv{e}_{12}/2} = \text{e}^{\theta \mv{e}_{12}} \vec{a} = \vec{a} \text{e}^{-\theta \mv{e}_{12}} ,
\eea
so that it is trivial to prove that the norm of vectors is preserved:
\bea
\op{R}(\vec{a})^2 = \op{R}(\vec{a}) \op{R}(\vec{a}) =  \vec{a} \text{e}^{-\theta \mv{e}_{12}} \text{e}^{\theta \mv{e}_{12}} \vec{a} = \vec{a}^2 .
\eea

The fact that a two-dimensional Euclidean vector and a complex number have the same number of components and the same signature is suggestive, and has often been exploited to turn two-dimensional problems into one-complex-dimensional problems.  Geometric algebra can perform this mapping directly by choosing an arbitrary unit vector $\vec{r}^2 = 1$ and multiplying it into all vectors:
\bea
\pv{a} \equiv \vec{a}\vec{r} = \grade{\vec{a}\vec{r}}_0 + \grade{\vec{a}\vec{r}}_2.
\eea
This special even multivector is called a \emph{paravector}, and it is easily seen that the norm of the original vector and the norm of the paravector are the same, regardless of the choice of $\vec{r}$:
\bea
\pv{a} \tilde{\pv{a}} = \vec{a} \vec{r}^2 \vec{a} = \vec{a}^2 .
\eea
Being an even multivector, this paravector transforms under three types of multiplication:
\begin{subequations}
\bea
\op{R}(\pv{a}) = \mv{U}\pv{a}\tilde{\mv{U}} \in \Cl^+_2 &\qquad& \text{Two-sided multiplication} \\
\op{R}_L(\pv{a}) = \mv{U}\pv{a} \in \Cl^+_2 &\qquad& \text{Left multiplication} \\
\op{R}_R(\pv{a}) = \pv{a}\mv{U} \in \Cl^+_2 &\qquad& \text{Right multiplication}
\eea
\end{subequations}
where $\mv{U}$ is a unitary even multivector.  The latter two cases correspond to \eq{spinordef}, which defines spinors; therefore, we associate paravectors $\pv{a} \in \Cl^+_2$ with spinors in two dimensions.  (Note that although spinors and vectors are simply related in two dimensions, this is not true in higher dimensions.)  Since objects in the even subalgebra are isomorphic to complex numbers, and the spin group is $\text{U}(1)$, calling complex numbers spinors is consistent.

Although spinors are trivial in two dimensions, the technique we use to obtain them is easily generalized to higher dimensions.  Before moving to the first interesting case of $\spin(3)=\SU(2)$ spinors, though, we will consider the anti-Euclidean case, which will be useful in later sections.

\subsection{The Quaternions: $\Cl_{0,2}\simeq\mathbb{H}$}
\label{quaternions}

A suitable basis for the anti-Euclidean geometric algebra $\Cl_{0,2}$ is
\bea
1,\ \mv{i}_1,\ \mv{i}_2,\ \mv{i}_3 \equiv \mv{i}_1 \mv{i}_2 ,
\eea
where the nonscalar basis elements obey Hamilton's equation for quaternions:
\bea
\mv{i}_1{}^2 = \mv{i}_2{}^2 = \mv{i}_3{}^2 = \mv{i}_1\mv{i}_2\mv{i}_3 = -1 ,
\label{hamilton}
\eea
which explains the notation $\mv{i}_3$ instead of $\mv{i}_{12}$.  Thus, we say that the quaternion algebra $\mathbb{H}$ is isomorphic to $\Cl_{0,2}$.  A vector in this space has a negative-definite norm, which is opposite in sign to the norm of the even multivector:
\bea
\vec{a}^2 < 0, \quad \grade{\mv{A}}_+ \widetilde{\grade{\mv{A}}}_+ > 0 .
\eea
The even subalgebra $\Cl^+_{0,2}\simeq \Cl_{0,1}$ is isomorphic to the even subalgebra for the Euclidean case $\Cl^+_2$, so that $\spin(0,2)\simeq\spin(2)=\text{U}(1)$ as expected, and the spinors will be the same.

However, there are other properties of quaternions that will prove to be useful.  Consider the most general multivector in $\Cl_{0,2}$, which is a full quaternion:
\bea
\mv{Q} = q_0 + q_k \mv{i}_k ;
\eea
the quaternion norm is a positive-definite scalar
\bea
\bar{\mv{Q}}\mv{Q} = q_0{}^2 + q_1{}^2 + q_2{}^2 + q_3{}^2 > 0,
\eea
where $\bar{\mv{Q}}$ is the Clifford conjugate.  This norm is invariant under $\text{U}(1)$, but it is also invariant under a more interesting transformation:
\bea
\mv{Q}' = \mv{\Theta}\mv{Q} \quad \text{where} \quad \mv{\Theta}\bar{\mv{\Theta}} = 1,
\eea
$\mv{\Theta}$ also being a quaternion.  In other words, quaternions can be regarded as spinors themselves under the action of the group consisting of unit quaternions.

Understanding how a quaternion can be isomorphic to a spinor requires looking back at the algebra $\Cl_{0,2}$.  A quaternion can be rewritten as
\bea
\mv{Q} = (q_0 + \mv{i}_3 q_3 ) + (q_2 - \mv{i}_3 q_1) \mv{i}_2 = \mv{Q}_1 + \tilde{\mv{Q}}_2 \mv{i}_2 ,
\label{quatasvec}
\eea
which behaves like a complex two-dimensional vector with $\mv{i}_3$ as the imaginary unit \cite{porteous}, and ``complex'' components $\mv{Q}_A$.  (The reverse in the second term looks strange, but it will allow us to make contact with the column vector view of spinors.)  Consider the Hermitian inner product, which is designed to preserve the complex structure while eliminating the vector parts:
\bea
\grade{\mv{Q},\mv{R}}_H \equiv \grade{\bar{\mv{Q}}\mv{R}}_+ = \frac{1}{2} \left( \bar{\mv{Q}}\mv{R} + \hat{\bar{\mv{Q}}}\hat{\mv{R}}\right) = \tilde{\mv{Q}}_1\mv{R}_1 + \tilde{\mv{Q}}_2 \mv{R}_2
\label{hermprod2}
\eea
which is the quadratic form preserved by the group $\SU(2)=\spin(3)$ \cite{porteous}.  We can rewrite this product in a more enlightening way (see Refs.~[\onlinecite{statesandops}] and [\onlinecite{lie_groups}]) by noting that $\hat{\mv{Q}} = - \mv{i}_3 \mv{Qi}_3$:
\bea
\grade{\mv{Q},\mv{R}}_H = \frac{1}{2}\left( \bar{\mv{Q}}\mv{R} - \mv{i}_3 \bar{\mv{Q}}\mv{R}\mv{i}_3 \right) = \bar{\mv{Q}}\mv{R} + \left\{ \left( \bar{\mv{Q}}\mv{R}\right) \times \mv{i}_3 \right\} \mv{i}_3,
\eea
where $2\mv{A}\times\mv{B}=\mv{AB}-\mv{BA}$ is the commutator product.  The reverse corresponds to the Hermitian conjugate, and affects the inner product in the expected way (recalling that $\hat{\bar{\mv{Q}}} = \tilde{\mv{Q}}$):
\bea
\widetilde{\grade{\mv{Q},\mv{R}}}_H = \grade{\mv{R},\mv{Q}}_H .
\label{hermprod2a}
\eea

Thus we see that the algebra $\Cl_{0,2} \simeq \mathbb{H}$ contains a natural complex and Hermitian structure, and as such can represent two-dimensional complex vectors.  This fact will be useful in the following section, when we treat Pauli spinors of the rotation group in three dimensions.

\section{Three Dimensions:  $\SU(2)$ Spinors}
\label{su2}

For the Euclidean three-dimensional algebra $\Cl_3$, we write the basis multivectors
\bea
1,\ \vec{\sigma}_1,\ \vec{\sigma}_2,\ \vec{\sigma}_3,\ I \vec{\sigma}_1 = \vec{\sigma}_2 \vec{\sigma}_3,\ I \vec{\sigma}_2 = \vec{\sigma}_3 \vec{\sigma}_1,\ I \vec{\sigma}_3 = \vec{\sigma}_1 \vec{\sigma}_2,\ I = \vec{\sigma}_1 \vec{\sigma}_2\vec{\sigma}_3
\eea
where $I^2 = -1$ is the pseudoscalar, which commutes with all elements of the algebra, the $\vec{\sigma}_i$ are vectors, and the $I \vec{\sigma}_i$ are bivectors.  The vectors square to unity $\vec{\sigma}_k{}^2 = 1$, and the bivectors have a negative norm $(I \vec{\sigma}_k)^2 = -1$.  Although the space of bivectors is three-dimensional, all bivectors are simple---they can be written as the outer product of two vectors.  This means that an element of the group $\spin(3)=\SU(2)$ can still be written as
\bea
\mv{U} = \text{e}^{\mv{i} \theta/2} = \cos (\theta/2) + \mv{i} \sin (\theta/2)
\eea
for some unit bivector $\mv{i}^2 = -1$ and scalar angle $\theta$.  (The minus sign from \eq{exponent} is not needed here, since for all bivectors $\mv{B}$ it is possible to write $-\exp(\mv{B}) = \exp(\mv{B}')$.)  However, in general two elements will not commute, $\mv{UU}' \neq \mv{U}'\mv{U}$, since in general their defining bivectors do not commute, $\mv{ii}' \neq \mv{i}'\mv{i}$.

As before, we can map any vector onto an element of the even subalgebra $\Cl^+_3 \simeq \Cl_{0,2}$ by selecting a unit vector $\vec{r}^2 = 1$:
\bea
\pv{a} = \vec{a}\vec{r} = \grade{\vec{a}\vec{r}}_0 + \grade{\vec{a}\vec{r}}_2;
\eea
the quadratic form is again restored by using the reverse
\bea
\pv{a}\tilde{\pv{a}} = \vec{a}\vec{r}^2 \vec{a} = \vec{a}^2 .
\label{pvnorm3}
\eea
For simplicity, let $\vec{r} = \vec{\sigma}_3$ so that
\bea
\pv{a} = a_3 + I \left( a_1 \vec{\sigma}_2 - a_2 \vec{\sigma}_1 \right) \equiv a_3 + a_1 \mv{i}_2 + a_2 \mv{i}_1,
\label{pv3}
\eea
where we have made the following identifications:
\bea
\mv{i}_1 = - I \vec{\sigma}_1,\ \mv{i}_2 = I \vec{\sigma}_2,\ \mv{i}_3 = I \vec{\sigma}_3 ,
\eea
which obey \eq{hamilton}.  The paravector in \eq{pv3} is the sum of a scalar and a bivector in $\Cl_3$, but we can also regard it as the sum of a scalar and a vector in $\Cl_{0,2}$, just as the paravectors in Section~\ref{u1} can be considered as ``complex numbers'' in $\Cl_{0,1}$.  By extension, the rotation multivector $\mv{U}$ is a unit quaternion in $\Cl_{0,2}$.

The reverse in $\Cl_{0,2}$ (which acts like the Hermitian conjugate, according to \eq{hermprod2a}) corresponds to an operation in $\Cl_3$ which we call the spin conjugate,
\bea
\mv{A}^\dagger = \vec{r} \tilde{\mv{A}} \vec{r} \qquad \to \quad \left( \mv{AB} \right)^\dagger = \mv{B}^\dagger \mv{A}^\dagger,
\eea
from which it is easy to see that a paravector is Hermitian:
\bea
\pv{a}^\dagger = \vec{r}\left(\vec{r}\vec{a}\right)\vec{r} = \pv{a}.
\eea
If we take Hermicity to be a defining characteristic of paravectors (in the same way that $\tilde{\vec{a}} = \vec{a}$ is a characteristic of vectors), the rotation operator takes a slightly different form:
\bea
\op{R}_{pv} (\pv{a}) = \mv{U}\pv{a}\mv{U}^\dagger ,
\eea
where $\mv{U}$ is the same even multivector as before.  \eq{pvnorm3} is preserved under this operation, as we can easily show:
\bea
\op{R}_{pv}(\pv{a}) \widetilde{\op{R}_{pv}(\pv{a})} = \mv{U}\pv{a}\mv{U}^\dagger \tilde{\mv{U}}^\dagger \tilde{\pv{a}} \tilde{\mv{U}} = \mv{U}\pv{a} \left( \tilde{\mv{U}}\mv{U}\right)^\dagger \tilde{\pv{a}} \tilde{\mv{U}} = \pv{a}\tilde{\pv{a}}.
\eea

All of this suggests that a paravector can be decomposed into two quaternions as
\bea
\pv{a} = \mv{\psi} \mv{\psi}^\dagger ;
\eea
since we know from Section~\ref{quaternions} that quaternions transform under left-sided multiplication by the unitary multivector $\mv{U}$, this is a \emph{spinor decomposition}, and we call the objects $\mv{\psi}$ \emph{Pauli spinors} \cite{realspinors,lounesto}.  (This is not the same decomposition as used by Cartan \cite{cartan_spin}, who complexifies the vector space so that he can assign spinors to null vectors---which amounts to using the sum of vectors and bivectors in $\Cl_3$.  We will use something closer to Cartan's approach in Section~\ref{lorentzspin}.)  The Hermitian product for Pauli spinors is
\bea
\grade{\mv{\psi},\mv{\phi}}_H = \grade{\mv{\phi},\mv{\psi}}^\dagger_H = \frac{1}{2} \left( \tilde{\mv{\psi}}\mv{\phi} + \mv{\psi}^\dagger \tilde{\mv{\phi}}^\dagger \right) = \tilde{\mv{\psi}}\mv{\phi} + \left\{ \vec{r} \times \left( \tilde{\mv{\psi}}\mv{\phi}\right) \right\} \vec{r} ;
\eea
since Pauli spinors are isomorphic to quaternions, we can use \eq{hermprod2} by analogy (with the reverse replacing the Clifford conjugate, and $\vec{r} = \vec{\sigma}_3$ for simplicity) to obtain
\bea
\grade{\mv{\psi},\mv{\phi}}_H = \tilde{\mv{\psi}}\mv{\phi} + \left\{ \left( \tilde{\mv{\psi}}\mv{\phi}\right) \times \mv{i}_3 \right\} \mv{i}_3 .
\label{hermprod3}
\eea

The Hermitian product finds the $\{1,\mv{i}_3\}$ components of $\tilde{\mv{\psi}}\mv{\phi}$; we extend this projection to any even multivector, which we write as
\bea
\grade{ \mv{\psi} }_C \equiv \mv{\psi} + \left( \mv{\psi} \times \mv{i}_3 \right) \mv{i}_3 = \frac{1}{2} \left( \mv{\psi} + \tilde{\mv{\psi}}^\dagger \right) \qquad \mv{\psi} \in \Cl^+_3 ,
\label{comproj3}
\eea
so that $\grade{\mv{\psi},\mv{\phi}}_H = \big\langle\tilde{\mv{\psi}}\mv{\phi}\big\rangle_C$.  We define the quaternion object
\bea
\mv{z} \equiv \mv{\psi} \big\langle\tilde{\mv{\psi}}\big\rangle_C = \frac{1}{2} \left( \mv{\psi}\tilde{\mv{\psi}} + \mv{\psi}\mv{\psi}^\dagger \right) \equiv \frac{1}{2} \left( \rho + \vec{s} \vec{\sigma}_3 \right)
\label{obs3}
\eea
that combines the (scalar) density
\bea
\rho \equiv \mv{\psi}\tilde{\mv{\psi}} ,
\eea
and the (vector) spin (see \emph{e.g.} Refs~[\onlinecite{STA,GA03}])
\bea
\vec{s} \equiv \mv{\psi}\vec{\sigma}_3 \tilde{\mv{\psi}} = \mv{\psi}\mv{\psi}^\dagger \vec{\sigma}_3 \qquad \to \quad \vec{s}^2 = \rho^2 .
\eea
The ordinary spin components are found by forming scalar products:
\bea
s_k = \grade{ \mv{\psi}\vec{\sigma}_3 \tilde{\mv{\psi}} \vec{\sigma}_k }_0 = \grade{ \vec{\sigma}_3 \tilde{\mv{\psi}} \vec{\sigma}_k \mv{\psi}}_0 .
\eea
The density and spin are the only independent measurable quantities obtainable from Pauli spinors, according to quantum theory.

We can invert \eq{obs3} formally to find any Pauli spinor in terms of the observables $\rho$ and $\vec{s}$:
\bea
\mv{\psi} = \mv{z} p^{-1} \mv{V}
\label{psiofz3}
\eea
where
\bea
p^2 = \big\langle\tilde{\mv{\psi}}\big\rangle_C \big\langle\mv{\psi}\big\rangle_C = \frac{1}{2} \left( \rho + \grade{ \vec{s} \vec{\sigma}_3 }_0 \right)
\eea
is a scalar and
\bea
\tilde{\mv{V}} = \mv{V}^\dagger = \mv{V}^{-1} \qquad \to \quad \mv{V} = \text{e}^{\mv{i}_3 \alpha}
\eea
is a unitary rotor with some phase angle $\alpha$.  Although $\mv{V}$ is formally dependent on $\tilde{\psi}$, it is arbitrary in that it does not affect the observables:
\bea
\mv{\psi}\tilde{\mv{\psi}} &=& \left( \mv{z} p^{-1} \text{e}^{\mv{i}_3 \alpha} \right) \left( \text{e}^{- \mv{i}_3 \alpha} p^{-1} \tilde{\mv{z}} \right) = p^{-2} \mv{z}\tilde{\mv{z}} = \rho ,\nonumber \\
\mv{\psi\psi}^\dagger &=& \left( \mv{z} p^{-1} \text{e}^{\mv{i}_3 \alpha} \right) \left( \text{e}^{- \mv{i}_3 \alpha} p^{-1} \mv{z}^\dagger \right) = p^{-2} \mv{zz}^\dagger = \vec{s\sigma}_3 
\eea
since
\bea
\tilde{\mv{z}} = \frac{1}{2} (\rho + \vec{\sigma}_3 \vec{s}) \qquad \text{and} \qquad \mv{z}^\dagger = \mv{z} .
\eea

Pauli spinors occur in nonrelativistic quantum theory, which is governed by the Schr\"{o}dinger-Pauli equation.  In the absence of interactions that couple to spin, the Schr\"{o}dinger-Pauli equation reduces to the ordinary Schr\"{o}dinger equation, and it is possible to interpret $\psi$ as a quantum state ``vector'' residing in a \emph{spin eigenstate} \cite{gurtler}.  Such an approach may help elucidate the classical field theory of the Schr\"{o}dinger equation.  Although we have never seen this done, using \eq{psiofz3} in the quantum mechanical equations of motion will give rise to a ``hydrodynamic'' formulation, by analogy with the Dirac case (see Section~\ref{bilinears}).

\section{Four-Dimensional Spacetime:  $\text{SL}(2,\mathbb{C})$ Spinors}
\label{sl2c}

\subsection{Dirac Spinors}

For spacetime, consider the Lorentz-signature geometric algebra $\Cl_{1,3}$ with the vector basis $\vec{\gamma}_\mu$:
\bea
\vec{\gamma}_0{}^2 = 1, \ \vec{\gamma}_1{}^2 = \vec{\gamma}_2{}^2 = \vec{\gamma}_3{}^2 = -1 .
\eea
(For the opposite spacetime signature, see Appendix~\ref{stsig}.)  The spin group $\spin_+(1,3)=\SL(2,\mathbb{C})$ is the double covering of $\SO_+(1,3)$, which is called the (proper, orthochronous) Lorentz group.  Elements of $\spin_+(1,3)$ can be written as
\bea
\mv{U} = \pm \text{e}^{\mv{B}/2} ,
\eea
where it is not generally possible to write the bivector $\mv{B}$ as the outer product of two vectors.  When $\mv{B}^2 = 0$, the operation is called a \emph{null rotation}; when $\mv{B}$ is timelike, the operation is a rotation, and when $\mv{B}$ is spacelike, the Lorentz transformation is a boost.

$\Cl_{1,3}$ contains six linearly-independent bivectors, which we write suggestively as
\bea
\vec{\sigma}_{k \neq 0} = \vec{\gamma}_k\vec{\gamma}_0 ,\ I \vec{\sigma}_k = \vec{\gamma}_{0123} \vec{\gamma}_k\vec{\gamma}_0 ,
\eea
since $\Cl^+_{1,3} \simeq \Cl_3$; we will also continue to use $I = \vec{\sigma}_{123} = \vec{\gamma}_{0123}$ to represent the pseudoscalar.  Paravectors are constructed using a timelike unit vector $\vec{t}^2 = 1$, so that
\bea
\pv{a} = \vec{at},
\eea
and the spinor conjugate is
\bea
\mv{A}^\dagger = \vec{t} \tilde{\mv{A}} \vec{t}.
\eea
This construction is also known as the space-time split \cite{GA03};  note that using a spacelike unit vector $\vec{r}^2 = -1$ is also possible, in which case we use the conjugate $\mv{A}^\ddagger \equiv - \vec{r} \tilde{\mv{A}} \vec{r}$.  As the pattern we are developing suggests, general even multivectors $\mv{\Psi} \in \Cl^+_{1,3}$ are spinors, which we associate with \emph{Dirac spinors}.

The combination of spinor conjugate and reverse suggests that we have an ``even-odd'' grading within $\Cl^+_{1,3}$, which we write as
\bea
\grade{\mv{A}}^{\pm} \equiv \frac{1}{2} \left( \mv{A} \pm \tilde{\mv{A}}^\dagger \right) \qquad \mv{A} \in \Cl^+_{1,3} ;
\label{relevenodd}
\eea
we will call these \emph{Pauli even} or \emph{Pauli odd} gradings, to avoid confusion.  Pauli spinors in $\Cl^+_{1,3}$ are Pauli-even multivectors, and we write a Dirac spinor as \cite{GA03}
\bea
\mv{\Psi} &=& \mv{\psi}_I + \mv{\psi}_{II} \vec{rt}
\eea
where each $\mv{\psi}$ is a Pauli spinor, represented in $\Cl_{1,3}$ (a scalar added to a spacelike bivector), while $\vec{r}^2 = -1$ is a spacelike unit vector such that $\vec{r}\cdot\vec{t} = 0$.  The conjugates are easily written in this form
\bea
\tilde{\mv{\Psi}} &=& \tilde{\mv{\psi}}_I - \vec{rt} \tilde{\mv{\psi}}_{II} \nonumber \\
\mv{\Psi}^\dagger &=& \tilde{\mv{\psi}}_I + \vec{rt} \tilde{\mv{\psi}}_{II} \\
\tilde{\mv{\Psi}}^\dagger &=& \mv{\psi}_I - \mv{\psi}_{II} \vec{rt} ,\nonumber
\eea
where we have used the fact that $\mv{\psi}^\dagger = \tilde{\mv{\psi}}$ for Pauli spinors.    The Lorentz-invariant product $\tilde{\Psi}\Psi$ is the sum of a scalar and a pseudoscalar, but it is not Hermitian under the spinor conjugate, so we define instead the Hermitian product
\bea
\grade{\mv{\Psi},\mv{\Phi}}_H = \grade{\mv{\Phi},\mv{\Psi}}^\dagger_H = \big\langle \tilde{\mv{\Psi}}\mv{\Phi} \big\rangle^+ + \left\{ (\vec{rt}) \times \big\langle \tilde{\mv{\Psi}}\mv{\Phi} \big\rangle^+ \right\} \vec{rt} ,
\label{hermprod4}
\eea
which is the same form as \eq{hermprod3}.  If we let $\vec{t} = \vec{\gamma}_0$, $\vec{r} = \vec{\gamma}_3$, and write out the Pauli spinors along the same lines as \eq{quatasvec}, the Dirac spinor begins to look like a four-component ``complex vector'':
\bea
\mv{\Psi} = \mv{\Psi}_1 + \mv{\Psi}^\dagger_2 \mv{i}_2 - \left( \mv{\Psi}_3 + \mv{\Psi}_4 \mv{i}_2 \right) I \mv{i}_3,
\eea
where again $\mv{i}_2 = I \vec{\sigma}_2 = \vec{\gamma}_1\vec{\gamma}_3$.  The quaternion element $\mv{i}_3 = I \vec{\sigma}_3 = \vec{\gamma}_2 \vec{\gamma}_1$ is associated with the spin current in quantum theory (see Section~\ref{bilinears}).  The spinor inner product may then be written in terms of these quaternion components as
\bea
\grade{\mv{\Psi},\mv{\Phi}}_H = \mv{\Psi}^\dagger_1 \mv{\Phi}_1 + \mv{\Psi}^\dagger_2 \mv{\Phi}_2 - \mv{\Psi}^\dagger_3 \mv{\Phi}_3 - \mv{\Psi}^\dagger_4 \mv{\Phi}_4 ,
\eea
which is preserved by the group $\text{U}(2,2)=\text{O}(4,2)$, the conformal group in spacetime (see Ref.~[\onlinecite{balachandran}], for example).  Discussions of the conformal group lead to \emph{twistors} \cite{penrose_rindler2,2spinors,crumeyrolle,GA03}, which lie beyond the scope of this paper.

\subsection{Observables:  Bilinear Covariants and Fierz Identities}
\label{bilinears}

Once again, the Hermitian product defines a map onto the subspace $\{1,I\vec{\sigma}_3\}$, so following \eq{comproj3} we define the map
\beann
\grade{\mv{\Psi}}_C \equiv \grade{\mv{\Psi}}^+ + \left( \vec{\sigma}_3 \times \grade{\mv{\Psi}}^+ \right) \vec{\sigma}_3 = \frac{1}{4} \left( \mv{\Psi} + \tilde{\mv{\Psi}}^\dagger - I \vec{\sigma}_3 \left\{ \mv{\Psi} + \tilde{\mv{\Psi}}^\dagger \right\} I \vec{\sigma}_3 \right) \qquad \mv{\Psi} \in \Cl^+_{1,3},
\eeann
which follows the path
\beann
\Cl^+_{1,3} \stackrel{\grade{\mv{\Psi}}^+}{\longrightarrow} \{1,I\vec{\sigma}_k\} \stackrel{\grade{\mv{\Psi}}_C}{\longrightarrow} \{1,I \vec{\sigma}_3\}
\eeann
The first map is relative to $\vec{t} = \vec{\gamma}_0$, while the second is relative to $\vec{\sigma}_3 = \vec{\gamma}_3 \vec{\gamma}_0$.

As with \eq{obs3}, we form the product \cite{statesandops}
\bea
\mv{Z} &\equiv& \mv{\Psi}\big\langle\tilde{\mv{\Psi}}\big\rangle_C = \frac{1}{4} \left( \mv{\Psi}\tilde{\mv{\Psi}} + \mv{\Psi\Psi}^\dagger - \mv{\Psi} I \vec{\sigma}_3 \tilde{\mv{\Psi}} I \vec{\sigma}_3 - \mv{\Psi} I \vec{\sigma}_3 \mv{\Psi}^\dagger I \vec{\sigma}_3 \right) \nonumber \\
&\equiv& \frac{1}{4} \left( \rho \text{e}^{I \beta} + \vec{J\gamma}_0 - \mv{S} I \vec{\sigma}_3 - \vec{K\gamma}_3 \right) = \frac{1}{4} \left( \rho \text{e}^{I \beta} + \pv{J} - \mv{S} I \vec{\sigma}_3 + \pv{K}\vec{\sigma}_3 \right)
\label{obs4}
\eea
where the objects $\{\rho,\beta,\vec{J},\mv{S},\vec{K}\}$ are the \emph{bilinear covariants} of Dirac theory, and the final expression is in terms only of elements in the even subalgebra.  In terms of the usual Dirac matrix bilinears (see Kaku\cite{kaku}, for example), we have
\bea
\begin{array}{lllll}
\text{Scalar} & \quad & \bar{\psi}\psi & \quad \to \quad & \rho \cos \beta = \big\langle \mv{\Psi}\tilde{\mv{\Psi}} \big\rangle_0 \\
\text{Vector} & \quad & \bar{\psi} \gamma^\mu \psi & \quad \to \quad & \vec{J} = \mv{\Psi}\vec{\gamma}_0\tilde{\mv{\Psi}} = \mv{\Psi\Psi}^\dagger \vec{\gamma}_0 = \pv{J}\vec{\gamma}_0 \\
\text{Bivector (or tensor)} & \quad & \bar{\psi} \frac{i}{2} [ \gamma^\mu, \gamma^\nu ] \psi & \quad \to \quad & \mv{S} = \mv{\Psi} I \vec{\sigma}_3 \tilde{\mv{\Psi}} \\
\text{Pseudovector} & \quad & \bar{\psi} \gamma^5 \gamma^\nu \psi & \quad \to \quad & I \vec{K} = \mv{\Psi} I \vec{\gamma}_3\tilde{\mv{\Psi}} = \mv{\Psi} I \vec{\sigma}_3 \mv{\Psi}^\dagger \vec{\gamma}_0 = I \pv{K}\vec{\gamma}_0 \\
\text{Pseudoscalar} & \quad & \bar{\psi} \gamma^5 \psi & \quad \to \quad & I \rho \sin \beta = \big\langle \mv{\Psi}\tilde{\mv{\Psi}} \big\rangle_4
\end{array}
\eea
The usual component forms of these expressions are found by taking inner products, so that for example
\bea
J^\mu = \left( \mv{\Psi}\vec{\gamma}_0\tilde{\mv{\Psi}}\right) \cdot \vec{\gamma}^\mu = \vec{\gamma}_0 \cdot \left( \tilde{\mv{\Psi}} \vec{\gamma}^\mu \mv{\Psi}\right) .
\eea
Since the bilinears are the same physical quantities in both mathematical forms, $\vec{J}$ is the (vector) current, $\mv{S}$ is the (bivector) angular momentum, and $\vec{K}$ is the (vector) spin current, while
\bea
\mv{\Psi}\tilde{\mv{\Psi}} = \rho \text{e}^{I \beta} = \cos \beta + I \sin \beta
\eea
encompasses the probability density $\rho$, with a (pseudoscalar) phase factor defined by the Yvon-Takabayasi angle\cite{takabayasi} $\beta$.

As with the Pauli spinor case, we may use \eq{obs4} to find $\mv{\Psi}$ in terms of its bilinear covariants:
\bea
\mv{\Psi} = \mv{Z} p^{-1} \mv{V}
\label{psiofz4}
\eea
where
\bea
p^2 = \big\langle\tilde{\mv{\Psi}}\big\rangle_C \big\langle\mv{\Psi}\big\rangle_C = \frac{1}{4} \left( \rho \cos \beta + \grade{ \vec{J} \vec{\gamma}_0}_0 - \grade{ \mv{S} I \vec{\sigma}_3 }_0 - \grade{ \vec{K} \vec{\gamma}_3 }_0 \right)
\eea
is a scalar, and 
\bea
\tilde{\mv{V}} = \mv{V}^\dagger = \mv{V}^{-1} \qquad \to \quad \mv{V} = \text{e}^{I \alpha^k \vec{\sigma}_k}
\eea
is an $\SU(2)$ rotor.  The density and current are unaffected by the value of $\alpha^k$, while the spin axis $\vec{\gamma}_3$ is rotated in space.  \eq{psiofz4} is analogous to expressions in Refs.~\cite{crawford85,crawford86,lounesto}, but its phase group is $\SU(2)$, not $\text{U}(1)$; in addition, it holds even when $\rho = 0$, in contrast to Equation~3.24 of Ref.~[\onlinecite{statesandops}].

At first glance, it would appear that $\mv{Z}$ has too many degrees of freedom to describe $\mv{\Psi}$, but the bilinear covariants are not independent of each other.  Forming all possible products of bilinears, we find that the current and spin current have the same length and are orthogonal to each other:
\bea
\begin{array}{l}
\vec{J}^2 = - \vec{K}^2 = \rho^2 \\ \vec{J}\vec{K} = - \vec{KJ} = - I \rho \text{e}^{-I \beta} \mv{S} \\ \grade{\vec{JK}}_0 = 0
\end{array}
\label{JandK}
\eea
while the products involving the spin are
\bea
\begin{array}{ll}
\vec{J} \mv{S} = I \rho \text{e}^{-I \beta} \vec{K} \qquad & \mv{S}\vec{J} = I \rho \text{e}^{I \beta} \vec{K} \\
\vec{K}\mv{S} = I \rho \text{e}^{-I \beta} \vec{J} \qquad & \mv{S}\vec{K} = I \rho \text{e}^{I \beta} \vec{J}
\end{array}\qquad \mv{S}^2 = - \rho^2 \text{e}^{2 I \beta}
\label{fierz}
\eea
All these relations together are called the \emph{Fierz identities}.  When $\rho \neq 0$, we discover from \eq{JandK} that
\bea
\mv{S} = I \rho^{-1} \text{e}^{I\beta} \vec{JK},
\eea
which combined with the expression $\vec{J}^2 = - \vec{K}^2 = \rho^2$ reduces the degrees of freedom to 8, the desired number.  Using Eqs.~(\ref{JandK}) and (\ref{fierz}), we find after some algebra that
\bea
\mv{Z}\tilde{\mv{Z}} = \frac{1}{4} \rho \text{e}^{I \beta} \left( \rho \cos \beta + \grade{\vec{J}\vec{\gamma}}_0 - \grade{ \mv{S} I \vec{\sigma}_3 }_0 - \grade{\vec{K}\vec{\gamma}}_3 \right) ,
\eea
so that $\mv{Z}$ is null if $\mv{\Psi}$ is.

If $\rho = 0$, we find
\bea
\begin{array}{l}
\vec{J}^2 = \vec{K}^2 = 0 \\
\vec{JK} = 0 \\
\mv{S}^2 = 0 \\
\vec{J}\mv{S} = \vec{K}\mv{S} = 0
\end{array}
\qquad \Longrightarrow \quad
\begin{array}{l} 
\vec{J}\ \text{and}\ \vec{K}\ \text{are both null}\\
\vec{J}\propto \vec{K} \\
\mv{S}\ \text{is made up of only two vectors, one of which is null}\\
\vec{J}\ \text{and}\ \vec{K}\ \text{are in the plane of}\ \mv{S}
\end{array}
\eea
so we let $\vec{K} = h \vec{J}$ and $\mv{S} = \vec{Js}$ for some scalar $h$ and spacelike vector $\vec{s}$ such that $\grade{\vec{J}\vec{s}}_0 = 0$ \cite{lounesto}.  Thus,
\bea
\mv{Z} = \frac{1}{4} \vec{J}\left( \vec{\gamma}_0 - \vec{s} I \vec{\sigma}_3 - h \vec{\gamma}_3 \right) = \frac{1}{4} \pv{J} \left( 1 + ( h - I \tilde{\pv{s}} ) \vec{\sigma}_3 \right)
\label{Z2}
\eea
where $\tilde{\pv{s}} = \vec{\gamma}_0\vec{s}$; all objects in the second expression are in the even subalgebra.  Since $\mv{\Psi\Psi}^\dagger = \pv{J}$, we have the relation
\bea
\mv{Z}\mv{Z}^\dagger = p^2 \pv{J},
\eea
which after some work yields the result \cite{lounesto}
\bea
h^2 = 1 + \vec{s}^2 \leq 1 
\label{hands}
\eea
(since $\vec{s}^2 \leq 0$).  The geometric interpretation of $h$ and $\vec{s}$ is obtained in Section~\ref{obslorentz}.

Substituting \eq{psiofz4} into the Dirac equation gives rise to the so-called hydrodynamic formulation of Dirac theory (see \emph{e.g.} Takabayasi\cite{takabayasi} and Crawford\cite{crawford86}).  We will not pursue this avenue here, but we believe that the use of geometric algebra helps clarify the classical field theory of Dirac spinors.

\section{Two-Component Relativistic Spinors}
\label{lorentzspin}

Dirac spinors are the most obvious spinors for $\spin_+(1,3) \simeq \text{SL}(2,\mathbb{C})$ from the point of view we have been taking so far, but their algebra admits a further reduction (just as the Dirac matrix representation of Lorentz transformations is a reducible representation of the Lorentz group).  The more primitive objects, which we will obtain via a projective split, are known as \emph{Lorentz} or \emph{2-spinors}.  These objects have found applications in analyzing spacetime geometry \cite{penrose_rindler1}, and are the foundation of the Newman-Penrose formulation of general relativity \cite{newman-penrose}.

As we discussed in the previous section, when the Dirac density vanishes (that is, $\mv{\Psi}\tilde{\mv{\Psi}} = 0$),
\bea
\vec{J}^2 = \pv{J}\tilde{\pv{J}} = 0,
\eea
which means that the paravector current can be written as 
\bea
\pv{J} = J_0 ( 1 + \mv{e}) \qquad J_0 = \grade{ \vec{J} \vec{\gamma}_0 }_0
\eea
where $\mv{e}$ is a timelike bivector such that $\mv{e}^2 = 1$.  From $\pv{J}$ we define the projection operator
\bea
\pv{l} = \frac{1}{2} ( 1 + \mv{e} ) = \frac{\pv{J}}{2 J_0} ,
\label{projector}
\eea
which has the following properties \cite{baylis}:
\bea
\begin{array}{lllllll}
\pv{l}^2 = \pv{l} &\quad& \text{idempotence} &\qquad&
\pv{l} + \tilde{\pv{l}} = 1 &\quad& \text{complementarity}\\
\pv{l}\tilde{\pv{l}} = 0 &\quad& \text{nullity} &\qquad&
\pv{l}^\dagger = \pv{l} &\quad& \text{Hermicity}
\end{array}
\label{idemproperties}
\eea

\begin{figure}[t]
	\includegraphics[scale=0.5]{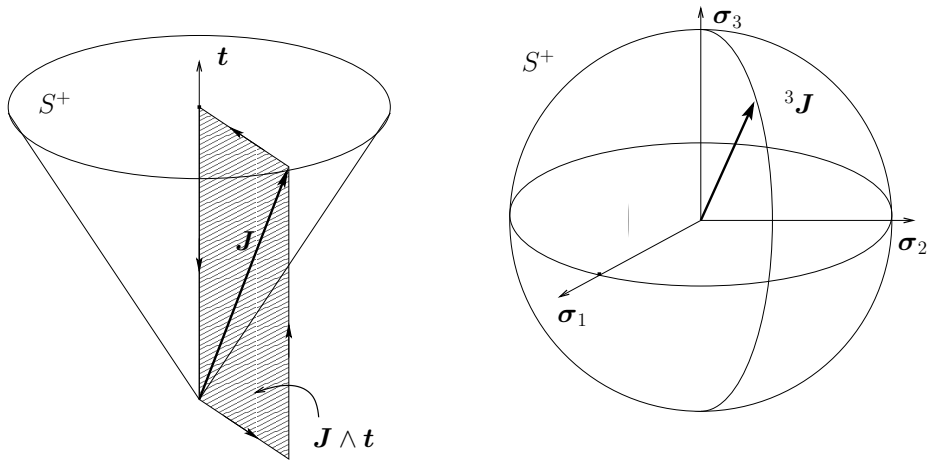} 
\caption{The future-directed light-cone, corresponding to a choice of timelike vector $\vec{t}$, maps a null vector $\vec{J}$ onto a three-dimensional vector $\vect{J}$, which is isomorphic to a bivector in $\Cl_{1,3}$: $\vect{J} = \vec{J}\wedge\vec{t}$.  The sphere that describes $\vect{J}$ and its spatial rotations is the anti-sky $S^+$.  Arbitrary spatial axes are given for reference.}
\label{skymap}
\end{figure}

A few words about geometry are in order:  since a null vector lies along the light-cone, and equivalently may be past- or future-directed, by selecting a timelike vector $\vec{t}$ and mapping $\vec{J}$ onto a paravector $\pv{J} = \vec{Jt}$, we restrict ourselves to either past or future.  The bivector part of the paravector can be seen as a vector $\vect{J}$ in three dimensions, whose length is the timelike (scalar) piece:
\bea
\pv{J} = \grade{\vec{Jt}}_0 + \grade{\vec{Jt}}_2 = J_t \left( 1 + \mv{e} \right) \qquad \to \qquad \pv{J}\tilde{\pv{J}} = 0 = J_t{}^2 - \left(\vect{J}\right)^2 ,
\eea
where $\vect{J} = J_t \mv{e}$.  If we factor out $J_t$, we have the description of a unit sphere in three dimensions, which plays much the same role as the light-cone:  any 3-vector lying inside the sphere represents the spatial piece of a timelike paravector, while any 3-vector that extends outside the sphere represents a spacelike paravector.  If $\vec{t}$ is past-directed, then the sphere is called the \emph{sky}, while if $\vec{t}$ is future-directed, the sphere is the \emph{anti-sky} \cite{penrose_rindler1} (see \fig{skymap}).

% Placeholder for figure sky.eps

\subsection{Lorentz Spinors and Flags}
\label{twospinors}

The projector in \eq{projector} can be used to split the even subalgebra into algebraic \emph{ideals}, which are spaces assigned to the sky and anti-sky.  Thus, any Dirac spinor can be written as
\bea
\mv{\Psi} = \mv{\Psi}\pv{l} + \mv{\Psi}\tilde{\pv{l}} \equiv \vec{\eta}_+ + \vec{\eta}_-
\label{skyantisky}
\eea
where the objects
\bea
\vec{\eta}_\pm \equiv \frac{1}{2} \mv{\Psi} ( 1 \pm \mv{e}) 
\label{lspinor}
\eea
are called \emph{Lorentz} or \emph{Infeld-van der Waerden spinors}, or simply \emph{2-spinors}.  In algebraic terms, these are known as \emph{half-} or \emph{semi-spinors} \cite{crumeyrolle}.  Since $\mv{\Psi}$ transforms under left multiplication by a unitary spinor, the quantity $\mv{\Psi}\pv{l}$ does as well, so 2-spinors are special cases of Dirac spinors.  However, to tie in with 2-spinor techniques, it is more interesting to build the space of Dirac spinors out of the 2-spinors, assuming the latter to be more primitive \cite{jones}.  In future calculations, unless otherwise noted, $\vec{\eta} \equiv \vec{\eta}_+$ for simplicity.

To justify our identification of $\mv{\Psi}\pv{l}$ with a two-component object, let us write the Dirac spinor as
\bea
\mv{\Psi} = \Psi_0 + \Psi_k \vec{\sigma}_k
\eea
where each $\Psi$ is the sum of a scalar and a pseudoscalar, and let $\mv{e} = \vec{\sigma}_3$.  These choices allow us to write
\bea
\vec{\eta} = \mv{\Psi}\pv{l} = \frac{1}{2} \left( \Psi_0 + \Psi_3 \right) ( 1 + \vec{\sigma}_3 ) + \frac{1}{2} \left( \Psi_1 + I \Psi_2 \right) \vec{\sigma}_1 ( 1 + \vec{\sigma}_3 ) \equiv \eta^A \vec{\alpha}_A
\eea
where the $\eta^A$ are scalars added to pseudoscalars, and the basis elements
\bea
\vec{o} = \vec{\alpha}_0 = \frac{1}{2} ( 1 + \vec{\sigma}_3 )\ , \quad \vec{\iota} = \vec{\alpha}_1 = \vec{\sigma}_1 \vec{o} = \text{e}^{-I \vec{\sigma}_2 \pi/2} \vec{o}
\label{spinbasis}
\eea
are rotations by $\pi$ of each other \cite{jones}.  In general, we will not assign $\mv{e}$, since $\mv{e} = \vec{\sigma}_3$ is a special case corresponding to Weyl spinors; see Section~\ref{obslorentz}.  Whatever the choice of $\mv{e}$, we have the identities
\beann
\vec{o}^\dagger = \vec{o} \qquad \text{and} \qquad \tilde{\vec{\iota}} = - \vec{\iota} .
\eeann

The reverse and spinor conjugate generate the dual and conjugate spinor spaces, which we write in the following abstract way:
\bea
\begin{array}{lll}
\vec{\eta} = \eta^A \vec{\alpha}_A \in \mathcal{S} & \qquad & \vec{\eta}^\dagger = \eta^{A'}{}^\dagger \vec{\alpha}^\dagger_{A'} \in \mathcal{S}^\dagger \\
\tilde{\vec{\eta}} = \eta^A \tilde{\vec{\alpha}}_A \in \tilde{\mathcal{S}} & \qquad & \tilde{\vec{\eta}}^\dagger = \eta^{A'}{}^\dagger \tilde{\vec{\alpha}}^\dagger_{A'} \in \tilde{\mathcal{S}}^\dagger
\end{array}
\eea
The spaces $\mathcal{S}$ and $\tilde{\mathcal{S}}^\dagger$ are both left ideals, while $\tilde{\mathcal{S}}$ and $\mathcal{S}^\dagger$ are right ideals; since they are all contained in the same geometric algebra, we can multiply spinors from different spaces without introducing any new products.  For example, for any two spinors $\vec{\eta}$ and $\vec{\xi}$
\bea
\tilde{\vec{\eta}}\vec{\xi} = - \tilde{\vec{\xi}}\vec{\eta} \qquad \text{and} \qquad \vec{\eta} \tilde{\vec{\xi}} = 0,
\label{symplectic_spinor}
\eea
which implies a symplectic structure.  If we define the following products in $\Cl^+_{1,3}$ (\emph{c.f.} Baylis\cite{baylis})
\bea
\grade{\mv{A,B}}_S &=& \grade{\mv{B,A}}_S \equiv \frac{1}{2} \left( \tilde{\mv{A}}\mv{B} + \tilde{\mv{B}}\mv{A} \right) = \grade{\tilde{\mv{A}} \mv{B}}_0 + \grade{\tilde{\mv{A}} \mv{B}}_4 \nonumber \\
\grade{\mv{A,B}}_B &=& - \grade{\mv{B,A}}_B \equiv \frac{1}{2} \left( \tilde{\mv{A}}\mv{B} - \tilde{\mv{B}}\mv{A} \right) = \grade{\tilde{\mv{A}}\mv{B}}_2 ,
\eea
we find that we have a representation of two-dimensional complex \emph{symplectic} geometric algebra \cite{crumeyrolle} embedded in $\Cl^+_{1,3}$, under the product $\tilde{\mv{A}}\mv{B}$.  The bivector element
\bea
\vec{\epsilon} = \grade{\vec{o},\vec{\iota}}_B = - \tilde{\vec{\epsilon}} \qquad \to \quad 2 \grade{\vec{\epsilon},\vec{\epsilon}^\dagger}_S = - 1
\eea
is the symplectic form, which allows us to write the spinor product as
\bea
\tilde{\vec{\eta}}\vec{\xi} = \left( \eta^0 \xi^1 - \eta^1 \xi^0 \right) \vec{\epsilon} .
\eea
By eliminating $\vec{\epsilon}$, we get the usual spinor inner product \cite{jones,lounesto86,benn}
\bea
\{ \vec{\eta},\vec{\xi} \} = - \{ \vec{\xi}, \vec{\eta} \} = - 2 \grade{ \tilde{\vec{\eta}}\vec{\xi}, \vec{\epsilon}^\dagger }_S = \eta^0 \xi^1 - \eta^1 \xi^0.
\eea
Although it does not behave quite like the Levi-Civita symbol, multiplication by $\vec{\epsilon}$ can be used to rotate the spinor basis elements:
\bea
\tilde{\vec{o}}^\dagger \vec{\epsilon} = \vec{\iota} \qquad \tilde{\vec{\iota}}^\dagger \vec{\epsilon} = - \vec{o} .
\label{spinormetric}
\eea
We can draw a more direct connection to component methods by rewriting dual spinors as
\bea
\tilde{\vec{\eta}} = \eta_A \vec{\alpha}^A,
\eea
so that $\{ \vec{\eta},\vec{\xi} \} = \eta_A \xi^A$.

\subsection{Flagpoles and Flags: Multivectors as Spinors}

The \emph{flagpole} of a spinor is its paravector current:
\bea
\pv{J} = \vec{\eta\eta}^\dagger \qquad \to \quad \pv{J}\tilde{\pv{J}} = \vec{\eta} ( \tilde{\vec{\eta}}\vec{\eta} )^\dagger \tilde{\vec{\eta}} = 0,
\eea
while the flag requires the introduction of a second spinor to define a spacelike paravector
\bea
\pv{L} = \vec{\eta\chi}^\dagger + \vec{\chi\eta}^\dagger \qquad \to \quad \tilde{\pv{L}}\pv{L} = - \{ \vec{\eta}, \vec{\chi} \}\{ \vec{\eta}, \vec{\chi} \}^\dagger  < 0 ,
\eea
from which the flag is found:
\bea
\mv{F} = \grade{ \pv{J},\pv{L} }_B = \tilde{\vec{\eta}}^\dagger \tilde{\vec{\eta}} \vec{\chi\eta}^\dagger = \{ \vec{\eta}, \vec{\chi} \} \tilde{\vec{\eta}}^\dagger \vec{\epsilon\eta}^\dagger
\qquad \to \quad \mv{F}^2 = \mv{F}\tilde{\mv{F}} = 0 .
\eea
(\eq{spinbasis} is helpful in evaluating these expressions.)  The flagplane is the set of all vectors in $\Cl_{1,3}$ of the form
\bea
\vec{\Pi} = ( a \pv{J} + b \pv{L} ) \vec{t},
\eea
where $a$ and $b > 0$ are (real) scalars.  The sum of two flagpoles yields a timelike paravector:
\bea
\pv{T} = \vec{\eta\eta}^\dagger + \vec{\chi\chi}^\dagger \qquad \to \quad \tilde{\pv{T}}\pv{T} = \{\vec{\eta},\vec{\chi}\} \{\vec{\eta},\vec{\chi}\}^\dagger > 0 .
\label{timelikepv}
\eea

% Placeholder for spinor component table
\begin{table}[t]
\begin{tabular}{|l||l|l|}\hline
\textbf{Object} & \textbf{GA Form} & \textbf{Component Form} \\ \hline
Null paravector (flagpole) & $\pv{J} = \vec{\eta\eta}^\dagger$ & $J^{AA'} = \eta^A\bar{\eta}^{A'}$ \\ \hline
Timelike paravector & $\pv{T} = \vec{\eta\eta}^\dagger + \vec{\tau\tau}^\dagger$ & $T^{AA'} = \eta^A\bar{\eta}^{A'} + \tau^A\bar{\tau}^{A'}$ \\ \hline
Spacelike paravector & $\pv{L} = \vec{\eta\chi}^\dagger + \vec{\chi\eta}^\dagger$ & $L^{AA'} = \eta^A\bar{\chi}^{A'} + \chi^A\bar{\eta}^{A'}$ \\ \hline
Null bivector (flag) & $\mv{F} = \tilde{\vec{\eta}}^\dagger \tilde{\vec{\eta}}\vec{\tau\eta}^\dagger$ & $F^{AA'BB'} = \eta^A \eta^B \epsilon^{A'B'} + \epsilon^{AB} \bar{\eta}^{A'} \bar{\eta}^{B'}$ \\ \hline
Null multivector & $\mv{N} = \vec{\eta\tau}^\dagger$ & $N^{AA'} = \eta^A \bar{\tau}^{A'}$ \\ \hline
Arbitrary bivector & $\mv{B} = \beta^{AB} \tilde{\vec{\alpha}}_A \vec{\epsilon} \vec{\alpha}_B$ & $B^{AA'BB'} = \beta^{AB}\epsilon^{A'B'} + \epsilon^{AB} \bar{\beta}^{A'B'}$ \\ \hline
Dirac spinor & $\mv{\Psi} = \vec{\eta} + \tilde{\vec{\chi}}^\dagger$ & $\Psi_\alpha = \eta_A + \chi^{A'}$ \quad (\emph{c.f.} Ref.~[\onlinecite{penrose_rindler1}])
\\ \hline
\end{tabular}
\caption{Geometric algebra and component forms for important objects constructed from 2-spinors.  For the right column, we use the overbar notation for complex conjugation \cite{penrose_rindler1}.  Note that $\epsilon^{AB} = - \epsilon^{BA}$ is the Levi-Civita symbol in two complex dimensions, while $\beta^{AB} = \beta^{BA}$ is a symmetric $2\times 2$ matrix consisting of scalars and pseudoscalars, such that $\tilde{\beta} = \beta$.}
\label{spinor_components}
\end{table}

In terms of components, the three types of paravectors are
\beann
\pv{J} &=& \eta^A \eta^{A'}{}^\dagger \vec{\alpha}_A \vec{\alpha}^\dagger_{A'} \qquad \pv{L} = \left( \eta^A \chi^{A'}{}^\dagger + \chi^A \eta^{A'}{}^\dagger \right) \vec{\alpha}_A \vec{\alpha}^\dagger_{A'} \qquad \pv{T} = \left( \eta^A \eta^{A'\dagger} + \chi^A \chi^{A'\dagger} \right) \vec{\alpha}_A \vec{\alpha}^\dagger_{A'},
\eeann
while the flag is
\bea
\mv{F} &=& \{ \vec{\eta}, \vec{\chi} \} \eta^{A'\dagger} \eta^{B'\dagger} \tilde{\vec{\alpha}}^\dagger_{A'} \vec{\epsilon} \vec{\alpha}^\dagger_{B'}.
\eea
Note that the quantity $\{ \vec{\eta}, \vec{\chi} \} \eta^{A'\dagger} \eta^{B'\dagger}$ contains all of the information of the ``real bivector'' described by Penrose and Rindler\cite{penrose_rindler1}, but the components are not real scalars---they are scalars added to pseudoscalars, which correspond to complex numbers in the ordinary treatment of spinors.  In addition, the null basis elements $\tilde{\vec{\alpha}}^\dagger_{A'} \vec{\epsilon} \vec{\alpha}^\dagger_{B'}$ can be evaluated using \eq{spinormetric}; the results are called a \emph{null tetrad}:
\bea
\tilde{\vec{\alpha}}^\dagger_{0} \vec{\epsilon} \vec{\alpha}^\dagger_{0} = \vec{\alpha}_1 \vec{\alpha}^\dagger_{0} \equiv \pv{m}^\dagger &\qquad& \tilde{\vec{\alpha}}^\dagger_{1} \vec{\epsilon} \vec{\alpha}^\dagger_{0} = -\vec{\alpha}_0 \vec{\alpha}^\dagger_{0} \equiv - \pv{l} \nonumber \\
\tilde{\vec{\alpha}}^\dagger_{0} \vec{\epsilon} \vec{\alpha}^\dagger_{1} = \vec{\alpha}_1 \vec{\alpha}^\dagger_{1} \equiv \pv{n} &\qquad& \tilde{\vec{\alpha}}^\dagger_{1} \vec{\epsilon} \vec{\alpha}^\dagger_{1} = - \vec{\alpha}_0 \vec{\alpha}^\dagger_{1} \equiv - \pv{m}
\eea
which spans the algebra $\Cl^+_{1,3}$.  Writing \[\pv{l} = \pv{E}_0\ , \quad \pv{n} = \pv{E}_1\ , \quad \pv{m} = \pv{E}_2\ , \ \text{and} \quad \pv{m}^\dagger = \pv{E}_3,\] we discover the Newman-Penrose metric for null tetrads:
\bea
\pv{E}_a \tilde{\pv{E}}_b + \pv{E}_b \tilde{\pv{E}}_a \equiv \eta_{ab} ,
\eea
where \cite{newman-penrose}
\bea
\left[ \eta_{ab} \right] \equiv \left( \begin{array}{cccc} 0 & 1 & 0 & 0 \\ 1 & 0 & 0 & 0 \\ 0 & 0 & 0 & -1 \\ 0 & 0 & -1 & 0 \end{array} \right) .
\eea

This development indicates that any real bivector can be written as
\bea
\mv{B} = \beta^{AB} \tilde{\vec{\alpha}}^\dagger_{A'} \vec{\epsilon} \vec{\alpha}^\dagger_{B'},
\eea
where $\beta^{AB} = \beta^{BA}$ is a matrix consisting of scalars and pseudoscalars; the symmetrization ensures that the basis elements span the bivectors only.  (This decomposition is of course not unique, since it depends on coordinates:  it is not possible in general to write a bivector simply in terms of 2-spinors, just as it is not always possible to decompose a bivector into two vectors.)  This means that all the information in a bivector can either be written in terms of a $4 \times 4$ real antisymmetric matrix, or a $2 \times 2$ matrix made up of scalars and pseudoscalars.  The latter is particularly useful in classifying the curvature tensor in general relativity\cite{fk03a}.  By extension, any null multivector (``complex null 4-vector''\cite{penrose_rindler1}) can be written as
\bea
\mv{N} = \vec{\eta\tau}^\dagger \qquad \to \quad \tilde{\mv{N}}\mv{N} = 0 .
\eea
As Table~\ref{spinor_components} shows, the component versions of two-spinor objects are arguably more unwieldy than their geometric algebra versions.  The tricky algebraic relations in standard texts are quite simple when using the symplectic structure of \eq{symplectic_spinor} to its greatest advantage.  (Our approach, which extends that of Jones and Baylis \cite{jones}, is likewise different from the real geometric algebra approach of Doran \emph{et al.} \cite{2spinors,GA03,statesandops}, especially with regard to the dual and conjugate spaces.)

A Dirac spinor is formed from the linear combination of spinors in the dual spaces \cite{2spinors}:
\bea
\mv{\Psi} = \vec{\eta} + \tilde{\vec{\chi}}^\dagger ;
\label{dirac2spin}
\eea
if we let
\bea
\vec{\eta} = \mv{\Phi}_1 \pv{l} \qquad \text{and} \qquad \vec{\chi} = \mv{\Phi}_2 \pv{l}
\eea
where $\mv{\Phi}_1$ and $\mv{\Phi}_2$ are Dirac spinors, we find in comparison with \eq{skyantisky} that
\bea
\mv{\Psi}\pv{l} = \vec{\eta} \qquad \text{and} \qquad \mv{\Psi}\tilde{\pv{l}} = \tilde{\vec{\chi}}^\dagger .
\eea
The paravector current is
\bea
\pv{J}_{Dirac} = \mv{\Psi\Psi}^\dagger = \vec{\eta\eta}^\dagger + \widetilde{ \vec{\chi\chi}^\dagger} ,
\eea
which is timelike, although it is not of the form of \eq{timelikepv}.

\subsection{Observables for Lorentz, Weyl, and Majorana Spinors}
\label{obslorentz}

Dirac spinors describe massive fermions, whereas Lorentz spinors are primarily used in discussions of spacetime geometry.  However, certain types of Lorentz spinors, called \emph{Weyl} spinors (used to describe massless fermions) and \emph{Majorana} spinors (massive fermions whose lepton number is not conserved); see \emph{e.g.} Kaku\cite{kaku}.  In the interests of knowing the relationship between these classes of spinor, it is useful to form their bilinears in the same manner as Section~\ref{bilinears}.

The observable properties of two-component spinors are found by substiting a Lorentz spinor into \eq{Z2}, so that it is identified with a singular Dirac spinor.  Rewriting \eq{lspinor} as
\bea
\vec{\eta} = \frac{1}{2} \mv{\Phi} ( 1 + \mv{e} ) \qquad \text{where} \quad \mv{e} \equiv \vec{r\gamma}_0
\eea
for some arbitrary Dirac spinor $\mv{\Phi}\tilde{\mv{\Phi}} \neq 0$ and spacelike vector $\vec{r}^2 = -1$, where $\grade{ \vec{r} \vec{\gamma}_0 }_0 = 0$, we can evaluate the bilinear covariants in terms of $\vec{r}$.  The current and spin current are easily found:
\bea
\pv{J} &=& \vec{\eta\eta}^\dagger = \frac{1}{2} \mv{\Phi} (1 + \mv{e}) \mv{\Phi}^\dagger = \frac{1}{2} \mv{\Phi} ( \vec{\gamma}_0 + \vec{r} ) \tilde{\mv{\Phi}} \vec{\gamma}_0 = \vec{J}\vec{\gamma}_0 \nonumber \\
\pv{K} &=& \vec{\eta\sigma}_3 \vec{\eta}^\dagger = \frac{1}{4} \mv{\Phi} (1 + \mv{e}) \vec{\sigma}_3 (1 + \mv{e}) \mv{\Phi}^\dagger = \frac{1}{2} \mv{\Phi} (1 + \mv{e}) \grade{\mv{e}\vec{\sigma}_3}_0  \nonumber \mv{\Phi}^\dagger = - \grade{ \vec{r} \vec{\gamma}_3 }_0 \pv{J} ,
\eea
where we have used the identities
\bea
\mv{e}\vec{\sigma}_3 = - \vec{\sigma}_3 \mv{e} + 2 \grade{\mv{e}\vec{\sigma}_3}_0 \qquad \text{and} \qquad \mv{e}\vec{\sigma}_3 = \vec{r}\vec{\gamma}_0 \vec{\gamma}_3 \vec{\gamma}_0 = - \vec{r} \vec{\gamma}_3 .
\eea
The angular momentum is
\bea
\mv{S} = \vec{\eta}I\vec{\sigma}_3 \tilde{\vec{\eta}} = \frac{1}{4} \mv{\Phi} (1 + \mv{e}) I \vec{\sigma}_3 (1 - \mv{e}) \tilde{\mv{\Phi}} = \frac{1}{2} \mv{\Phi} (1 + \mv{e}) \left(\mv{e} \times (I\vec{\sigma}_3 )\right) \tilde{\mv{\Phi}} ;
\eea
writing
\bea
1 = \big( \tilde{\mv{\Phi}}\mv{\Phi}\big)^\dagger (\omega^\dagger)^{-1} = \mv{\Phi}^\dagger \tilde{\mv{\Phi}}^\dagger (\omega^\dagger)^{-1} \qquad \text{where} \quad \omega \equiv \tilde{\mv{\Phi}}\mv{\Phi} ,
\eea
and defining
\bea
\mv{q} = - \mv{e} \times (I\vec{\sigma}_3 ) = I \grade{ \vec{r} \vec{\gamma}_3 }_2 ,
\eea
we find
\bea
\mv{S} = \frac{1}{2} \mv{\Phi} (1 + \mv{e})\left\{ \mv{\Phi}^\dagger \tilde{\mv{\Phi}}^\dagger (\omega^\dagger)^{-1} \right\} \left(\mv{e} \times (I\vec{\sigma}_3 )\right) \tilde{\mv{\Phi}} = \pv{J} \tilde{\pv{s}} = \vec{J} \vec{s}
\eea
where
\bea
\pv{s} \equiv (\omega^\dagger)^{-1} \mv{\Phi} \mv{q} \mv{\Phi}^\dagger = \vec{s} \vec{\gamma}_0 .
\eea
The norm of $\pv{s}$ is independent of $\mv{\Phi}$:
\bea
\pv{s}\tilde{\pv{s}} = (\omega^\dagger)^{-2}\left( \mv{\Phi} \mv{q} \mv{\Phi}^\dagger \right) \left( \tilde{\mv{\Phi}}^\dagger \tilde{\mv{q}} \tilde{\mv{\Phi}} \right) = - (\omega^\dagger)^{-1} \mv{\Phi} \mv{q}^2 \tilde{\mv{\Phi}} = - \mv{q}^2 = \grade{\mv{e}\vec{\sigma}_3}^2_0 - 1 ,
\eea
which agrees with \eq{hands} with the identification $h = \grade{\mv{e}\vec{\sigma}_3}_0 = - \grade{ \vec{r} \vec{\gamma}_3}_0$.

Now we are able to provide a geometrical interpretation for the objects $h$ and $\vec{s}$ from \eq{Z2}:
\bea
\vec{\gamma}_3 \vec{r} = \text{e}^{\mv{j}\varphi}
\eea
where
\bea
\mv{j} \sin \varphi = \grade{ \vec{\gamma}_3 \vec{r}}_2 ,
\eea
so that
\bea
\varphi = \tan^{-1} \frac{\sqrt{- \grade{ \vec{\gamma}_3 \wedge \vec{r}}_2{}^2}}{\grade{\vec{r} \vec{\gamma}_3}_0 } .
\eea
In plain language, the angle $\varphi$ determines the relative orientation of the projective vector $\vec{r}$ with respect to the spin axis $\vec{\gamma}_3$.  Thus, we are able to write
\bea
h = \cos \varphi \qquad \text{and} \qquad \vec{s} = \vec{c} \sin \varphi
\eea
for some spacelike vector $\vec{c}^2 = -1$.

For a general Lorentz spinor, $\mv{S}$ and $\vec{K}$ are nonvanishing quantities.  However, two special cases interest us:  $\vec{r} = \pm \vec{\gamma}_3$, and $\vec{r} = \pm \vec{\gamma}_2$.  The former case corresponds to \emph{Weyl spinors}:
\bea
\vec{\eta}^{W}_\pm = \frac{1}{2} \mv{\Phi} ( 1 \pm \vec{\sigma}_3 ),
\eea
which are eigenspinors of the \emph{chirality operator} (see Ref.~[\onlinecite{GA03}] for a derivation of this operator):
\bea
\op{\chi}_\pm (\vec{\eta}^{W}_\pm) = \frac{1}{2} \vec{\eta}^{W}_\pm ( 1 \pm \vec{\sigma}_3 ) = \vec{\eta}^{W}_\pm .
\eea
The projection angle $\varphi = 0$ or $\pi$, so that $\mv{S} = 0$ and
\bea
\mv{Z}_{W} = \frac{1}{4} \pv{J} \left( 1 \pm \vec{\sigma}_3 \right).
\eea
When $\vec{r} = \pm \vec{\gamma}_2$, we have a \emph{Majorana spinor}, 
\bea
\vec{\eta}^M_\pm = \frac{1}{2} \mv{\Phi} ( 1 \pm \vec{\sigma}_2 )
\eea
which is an eigenspinor of the charge conjugation operator:
\bea
\op{\mathcal{C}} (\vec{\eta}^M_\pm) = \vec{\eta}^M_\pm \vec{\sigma}_2 = \pm \vec{\eta}^M_\pm .
\eea
For this case, $\varphi = \pm \pi/2$, so that $\vec{K} = 0$ and
\bea
\mv{Z}_M = \frac{1}{4} \pv{J} \left( 1 + I \mv{s}\vec{\sigma}_3 \right),
\eea
where $\mv{s}^2 = 1$ is a bivector.

Majorana spinors are a special case of a general type defined by
\bea
\vec{r} = \cos \delta \vec{\gamma}_1 + \sin \delta \vec{\gamma}_2 \qquad \to \quad \grade{ \vec{r} \vec{\gamma}_3 }_0  = 0 ,
\eea
which Lounesto \cite{lounesto} calls \emph{flag-pole} spinors, since they are characterized by a null paravector current (flagpole) and null bivector angular momentum (flag):
\bea
\mv{Z}_\perp = \frac{1}{4} \pv{J} \left( 1 + I \mv{s}\vec{\sigma}_3 \right),
\eea
where again $\mv{s}^2 = 1$.  (When neither $\mv{S}$ nor $\pv{K}$ vanish, Lounesto calls the general 2-spinors \emph{flag-dipole} spinors.)

\section{Complex Algebras and Ideals of Algebras}
\label{ideals}

Typically a Pauli or Dirac spinor is regarded as a 2- or 4-dimensional complex column vector (from a physicist's standpoint), or the left-minimal ideal of a real or complex geometric algebra (from the mathematician's point of view).  The alternative view we take is due to Hestenes \cite{realspinors}; here we relate the three perspectives.

Consider the simple simpler case of Pauli spinors:  in Section~\ref{quaternions}, we already related the quaternions to complex 2-dimensional vectors by using the anti-Euclidean geometric algebra $\Cl_{0,2}$.  The imaginary unit $\mv{i}_3$ was naturally chosen in this algebra; when we start from $\Cl_3$, the imaginary unit (and hence ``complex conjugation'') depends on the choice of unit vector $\vec{r}$ we use to map vectors onto paravectors.  This is analogous to picking a spin axis in quantum mechanics---it is a necessary, yet arbitrary, choice, and selecting a different $\vec{r}$ will give a different $\mv{i}_3$.  A real geometric algebra may have multiple imaginary units, which have specific geometric interpretations, unlike the uninterpreted imaginary $i$.

It is possible to relate our spinor procedure to the more standard expositions by constructing the two-dimensional complex geometric algebra $\Ccl_2$.  Introducing a commutative imaginary unit to $\Cl_2$ doubles the size of the algebra:
\bea
1,\ i,\ \vec{e}_1,\ i \vec{e}_1,\ \vec{e}_2,\ i \vec{e}_2,\ \mv{e}_{12},\ i \mv{e}_{12} .
\eea
This algebra, however, is isomorphic to $\Cl_3$ if we identify $i$ with the pseudoscalar $I$ and let $\vec{\sigma}_3 = i \mv{e}_{12}$ be the third basis vector.  Another way to show this equivalence is through matrix representations (see Refs.~[\onlinecite{porteous,benn,crumeyrolle}]); we discuss representations of real geometric algebras in Appendix~\ref{matrep}.  Further work shows that it is always possible to represent even-dimensional complex geometric algebras in higher-dimensional real geometric algebras \cite{complex_algebra,hest86}.

Stating all of this still does not prove that our geometric construction is mathematically equivalent to the column-vector or left-ideal form.  The ideal structure \cite{crumeyrolle,benn}, for a judicious choice of matrix representation, is in fact equivalent to the column-vector form; both be obtained from the quaternion expression by multiplication:
\bea
\mv{\Psi} = \frac{1}{2} \mv{\psi} (1 + \vec{r})
\eea
where $\vec{r}^2 = 1$ is a unit vector in $\Cl_3$ \cite{lounesto}.  (That $\mv{\Psi}$ can be written as a complex column vector is best seen by using the Pauli matrices for $\vec{\sigma}_k$, and letting $\vec{r} = \vec{\sigma}_3$; see Appendix~\ref{matrep}.)  The object
\bea
\mv{l} = \frac{1}{2} (1 + \vec{r})
\eea
is an idempotent $\mv{l}^2 = \mv{l}$, and serves to project the quaternion spinor onto a space that is invariant under left multiplication by any object in the algebra $\Cl_3\simeq\Ccl_2$---and hence defines a left minimal ideal.

Our objection to the ideal construction is that the projector does not lie in the algebra of the group $\spin(3)$, and hence needs additional structure beyond that which is provided by the paravector space $\Cl^+_3\simeq\Cl_{0,2}$.  In fact, the left-ideal version does not add any new information, since the quaternion and ideal representations both have four real components (two complex components).  The geometric interpretation of a scalar added to a bivector is arguably simpler before the idempotent $\mv{l}$ is applied.

Dirac spinors may be treated in a similar way.  However, the complex 4-dimensional geometric algebra is isomorphic to a \emph{5-dimensional} real algebra:  $\Ccl_4 \simeq \Cl_{4,1} \simeq \Cl_{2,3} \simeq \Cl_5$.  The process of breaking the larger algebra down to $\Cl_{1,3}$ is done by Hestenes\cite{hest86}, while the equivalence of spinors in the complexified geometric algebra $\mathbb{C}\otimes\Cl_{1,3}$ to multivectors in $\Cl^+_{1,3}$ is shown by Lounesto\cite{lounesto}.

More on complex algebras in the context of matrix representations can be found in Appendix~\ref{matrep}.

\section{Generalization and Discussion}
\label{discussion}

We have explicitly constructed the spinors associated with the spin groups $\spin(2)$, $\spin(3)$, and $\spin_+(1,3)$, but the procedure here is completely general, and is readily extended.  Section~\ref{spingroups} laid out the algebra of spin groups in any dimension and signature, so that
\beann
\spin(p,q) &=& \left\{ \mv{U} \in \Cl^+_{p,q} \ \left| \ \mv{U}\tilde{\mv{U}} = \pm 1 \right. \right\} \\
\spin_+(p,q) &=& \left\{ \mv{U} \in \Cl^+_{p,q} \ \left| \ \mv{U}\tilde{\mv{U}} = 1 \right. \right\}
\eeann
An even multivector in $\Cl^+_{p,q}$ transforms under left- or right-multiplication by a member of the spin group, so we identify general even multivectors with spinors, and form the spinor space:
\bea
\Sigma_{p,q} = \left\{ \mv{\Psi} \in \Cl^+_{p,q} \ \left| \ \forall \mv{U} \in \spin(p,q)\ , \ \mv{U} \mv{\Psi} \in \Cl^+_{p,q} \right.\right\} .
\eea
The complex structure of spinor space is defined by some vector $\vec{r}$ in $\Cl_{p,q}$, so that the spinor conjugate is
\bea
\mv{A}^\dagger = \vec{r} \tilde{\mv{A}} \vec{r}^{-1} .
\eea

If some objects in $\Cl^+_{p,q}$ square to unity, then idempotent elements, which obey \eq{idemproperties}, can be constructed:
\bea
\pv{P}_\pm = \frac{1 \pm \mv{e}}{2} \qquad \forall \mv{e}\in \Cl^+_{p,q} \ , \ \mv{e}^2 = 1 .
\eea
\emph{Semi-spinors} are then formed by projecting the full spinor space across the idempotents:
\bea
\mv{\Psi} = \mv{\Psi} \pv{P}_+ + \mv{\Psi} \pv{P}_- = \vec{\eta}_+ + \vec{\eta}_-;
\eea
the $\vec{\eta}_\pm$ also are spinors, which belong to an irreducible represention of the group.  (Conversely, if idempotent elements exist, full spinors belong to a reducible representation of the spin group.)  The semi-spinor space is symplectic, since $\tilde{\vec{\eta}} \vec{\xi} = - \tilde{\vec{\xi}}\vec{\eta}$ for any two semi-spinors.  Thus, the inner product of semi-spinors is antisymmetric and does not necessarily yield ``complex'' scalar quantities, and extracting the scalar value of the product is basis-dependent, even though the product $\tilde{\vec{\eta}} \vec{\xi}$ is not.

Although the physical meaning of observables in a higher-dimensional theory would depend on the parameters of that theory, the method of finding observables used in Sections~\ref{su2}, \ref{bilinears}, and \ref{obslorentz} is generalizable to any dimensionality and signature.  As the calculation of the Fierz identities shows, the relationship between observables is easily found.  Equations (\ref{psiofz3}) and (\ref{psiofz4}) demonstrate a straightforward procedure to write spinors in terms of observables, in a way that improves upon the standard matrix-based methods.

Treating spinors in the context of a real geometric algebra provides a simplification of both mathematical method and physical interpretation.  Starting from group theory, we have reproduced all of the basic algebraic and geometric results of spinor theory, while avoiding the introduction of a complex unit $i$ whose physical geometric interpretation is unclear. We advocate this approach to spinors as the most straightforward and physically transparent technique available.

\begin{appendix}

\section{Matrix Representations of Geometric Algebras}
\label{matrep}

Geometric algebras may be represented as matrix algebras, which is how we exploit the analogue between the algebra of Pauli matrices and $\Cl_3$, for example.  Matrices whose entries are in the division rings of the real numbers $\mathbb{R}$, complex numbers $\mathbb{C}$, and quaternions $\mathbb{H}$ are isomorphic to real geometric algebras, which we catalogue here.

The real $2 \times 2$ matrices are spanned by the basis
\bea
\mv{1} = \left( \begin{array}{cc} 1 & 0 \\ 0 & 1 \end{array} \right)\ , \ \vec{e}_1 = \left( \begin{array}{cc} 1 & 0 \\ 0 & -1 \end{array} \right)\ , \ \vec{e}_2 = \left( \begin{array}{cc} 0 & 1 \\ 1 & 0 \end{array} \right)\ , \ \mv{e}_{12} = \vec{e}_1 \vec{e}_2 = \left( \begin{array}{cc} 0 & 1 \\ -1 & 0 \end{array} \right) ,
\eea
which we label for consistency with the treatment of Section~\ref{cl2}.  Denoting this algebra by $\mathbb{R}(2)$, we can easily show that $\mathbb{R}(2) \simeq \Cl_2$.  Also, if $\mv{e}_{12}$ is regarded as a vector, the algebra isomorphism is $\mathbb{R}(2) \simeq \Cl_{1,1}$.  We have already shown that $\Cl_{0,2} \simeq \mathbb{H}$, which can be represented by $2\times 2$ matrices with complex entries.  In a similar way, the Pauli matrices
\bea
\vec{\sigma}_1 = \left( \begin{array}{cc} 0 & 1 \\ 1 & 0 \end{array} \right)\ , \ \vec{\sigma}_2 = \left( \begin{array}{cc} 0 & -i \\ i & 0 \end{array} \right)\ ,\ \vec{\sigma}_3 = \left( \begin{array}{cc} 1 & 0 \\ 0 & -1 \end{array}\right)
\eea
behave like a vector basis for $\Cl_3$.  The particular idempotent element in $\Cl^+_{1,3} \simeq \Cl_3$
\bea
\pv{l} = \frac{1 + \vec{\sigma}_3}{2} = \left( \begin{array}{cc} 1 & 0 \\ 0 & 0 \end{array} \right)
\eea
shows why complex column spinors work:  letting $\mv{\Psi} = \Psi_0 \mv{1} + \Psi_k \vec{\sigma}_k$ be a general multivector (with $\Psi_\mu$ being complex numbers), we have
\bea
\vec{\eta} = \mv{\Psi}\pv{l} = \left( \begin{array}{cc} \Psi_0 + \Psi_3 & 0 \\ \Psi_1 + i \Psi_2 & 0 \end{array} \right) ,
\eea
which algebraically behaves like a column vector.

Rather than writing down all the matrix bases, we simply list the lower-dimensional algebras in terms of the total dimension $n = p + q$ and signature $s = p - q$ in Table~\ref{realalgebras}.  Several of these algebras are denoted as ${}^2\mathbb{F}(n)$; this is a weakness of matrix representations, since not all geometric algebras can be represented by square matrices.  Consider the 1-dimensional Euclidean algebra $\Cl_1 \simeq {}^2\mathbb{R}$; this is the direct sum of the algebra of real numbers with itself, which we write as
\bea
\mv{A} = ( a_1\ ,\ a_2 ) \in \mathbb{R} \oplus \mathbb{R} \qquad \to \quad \mv{AB} = (a_1 b_1 + a_2 b_2\ ,\ a_1 b_2 + a_2 b_1)
\eea
The geometric algebra way of writing these expressions is more straightforward, in the same manner as using $i$ in complex algebra:
\bea
\mv{A} = a_1 + a_2 \vec{e} \in \Cl_1 \qquad \to \quad \mv{AB} = a_1 b_1 + a_2 b_2 + ( a_1 b_2 + a_2 b_1)\vec{e} .
\eea
The matrix representation ${}^2\mathbb{R}(2)$, for example, then consists of $2 \times 2$ matrices whose elements are in $\mathbb{R} \oplus \mathbb{R}$.

% Placeholder for table of matrix representations
\begin{table}[t]
\begin{tabular}{| r | ccccccccccccccc |}\hline
%\multicolumn{16}{|c|} s = p - q \\ \hline
& s = -7 & -6 & -5 & -4 & -3 & -2 & -1 & 0 & 1 & 2 & 3 & 4 & 5 & 6 & 7 \\ \hline
\ n = 0\ &    &    &    &    &    &    &    & $\mathbb{R}$ & & & & & & & \\
1 &    &    &    &    &    &    & $\mathbb{C}$ & & ${}^2\mathbb{R}$ & & & & & & \\
2 &    &    &    &    &    & $\mathbb{H}$ & & $\mathbb{R}(2)$ & & $\mathbb{R}(2)$ & & & & & \\
3 &    &    &    &    & ${}^2\mathbb{H}$ & & $\mathbb{C}(2)$ & & ${}^2\mathbb{R}(2)$ & & $\mathbb{C}(2)$ & & & & \\
4 &    &    &    & $\mathbb{H}(2)$ & & $\mathbb{H}(2)$ & & $\mathbb{R}(4)$ & & $\mathbb{R}(4)$ & & $\mathbb{H}(2)$ & & & \\
5 &    &    & $\mathbb{C}(4)$ & & ${}^2\mathbb{H}(2)$ & & $\mathbb{C}(4)$ & & ${}^2\mathbb{R}(4)$ & & $\mathbb{C}(4)$ & & ${}^2\mathbb{H}(2)$ & & \\
6 &    & $\mathbb{R}(8)$ & & $\mathbb{H}(4)$ & & $\mathbb{H}(4)$ & & $\mathbb{R}(8)$ & & $\mathbb{R}(8)$ & & $\mathbb{H}(4)$ & & $\mathbb{H}(4)$ & \\
7 & ${}^2\mathbb{R}(8)$ & & $\mathbb{C}(8)$ & & ${}^2\mathbb{H}(4)$ & & $\mathbb{C}(8)$ & & ${}^2\mathbb{R}(8)$ & & $\mathbb{C}(8)$ & & ${}^2\mathbb{H}(4)$ & & $\mathbb{C}(8)$ \\ \hline
\end{tabular}
\caption{Matrix representations for $\Cl_{p,q}$, $p+q > 8$; as before, $n = p + q$ and $s = p - q$.  $\mathbb{F}(n)$ stands for an $n\times n$ matrix in the field $\mathbb{F} = \{\mathbb{R},{}^2\mathbb{R},\mathbb{C},\mathbb{H},{}^2\mathbb{H}\}$.  (This table is adapted from Lounesto\cite{lounesto}.)}
\label{realalgebras}
\end{table}

Complex geometric algebras $\Ccl_n$ are all of positive signature, so the matrix representation depends only on dimension.  The even-dimensional algebras are the most useful for spinors ($\SU(3)$ gauge theory, used in quantum chromodynamics, does not use spinors since $\SU(3)$ is not a spin group.)  The matrix representations for the even-dimensional algebras are
\bea
\Ccl_{2 k} \simeq \mathbb{C}(2^k),
\eea
so that $\Ccl_0 \simeq \mathbb{C} \simeq \Cl_{0,1}$, $\Ccl_2 \simeq \mathbb{C}(2) \simeq \Cl_3$, $\Ccl_4 \simeq \mathbb{C}(4) \simeq \Cl_{4,1}$, and so forth.  The odd-dimensional complex algebras have matrix representations of the form
\bea
\Ccl_{2 k + 1} \simeq {}^2\mathbb{C}(2^k) ,
\eea
so that $\Ccl_1 \simeq {}^2\mathbb{C}$, $\Ccl_3 \simeq {}^2\mathbb{C}(2)$, \emph{etc.}  These algebras are not isomorphic to real geometric algebras, and must be embedded in real algebras of higher dimension to involve commuting pseudoscalar elements.

\section{A Note About Spacetime Signature}
\label{stsig}

In this paper, along with the papers of Hestenes, Doran, Lasenby, and Gull, the spacetime signature of trace $-2$ is used (the ``west coast'' metric), which corresponds to the geometric algebra $\Cl_{1,3}$.  Frequently, the opposite signature is used, which involves the algebra $\Cl_{3,1}$:
\bea
\vec{\lambda}_0{}^2 = -1\ ,\ \vec{\lambda}_1{}^2 = \vec{\lambda}_2{}^2 = \vec{\lambda}_0{}^3 = 1 .
\eea
As we discussed in Section~\ref{spingroups}, the even subalgebras of these two spaces are isomorphic---$\Cl^+_{1,3} \simeq \Cl^+_{3,1}$---which means the spinors associated with the two spaces are the same.  The paravectors, however, will maintain the signature of the originating algebra, due to the way they are defined.

A compelling reason to consider $\Cl_{1,3}$ over $\Cl_{3,1}$ is motivated by the isomorphism \[\Cl_{1,3} \simeq \Cl_4 ,\] whereas $\Cl_{3,1} \simeq \Cl_{2,2}$.  Frequently in quantum field theory, it is useful to transform from a space of Lorentz signature to a space with Euclidean signature, which allows the evaluation of correlation functions.  The \emph{Wick rotation} in geometric algebra is
\bea
\op{W}:\ \Cl_{1,3} \to \Cl_4 \qquad \op{W} (\vec{a}) = \grade{ \vec{a} \vec{\gamma}_0}_0 \vec{\gamma}_0 + \grade{ \vec{a} \vec{\gamma}_0 }_2 = \cvec{a},
\eea
where $\vec{a} \in \grade{\Cl_{1,3}}_1$, and $\cvec{a}$, which is the sum of a timelike vector and a timelike bivector in $\Cl_{1,3}$, is interpreted as being a vector in $\Cl_4$.  Thus, we have mapped the basis elements as follows:
\bea
\op{W}(\vec{\gamma}_0 ) = \vec{\gamma}_0\ , \ \op{W}(\vec{\gamma}_{k\neq 0} ) = \vec{\gamma}_k\vec{\gamma}_0 = \vec{\sigma}_k .
\eea
To work with the Wick-rotated multivectors, the scalar product (for example) must be considered in $\Cl_4$ instead of $\Cl_{1,3}$, so we define
\bea
\grade{ \cvec{a} \cvec{b}}_0 = \frac{1}{2} \left( \cvec{a} \cvec{b} + \cvec{b} \cvec{a} \right) = a_0 b_0 + a_1 b_1 + a_2 b_2 + a_3 b_3 .
\eea
A Wick rotation to a Euclidean space from $\Cl_{3,1}$ is typically done by using a unit imaginary, which we do not need for $\Cl_{1,3}$.

%\section{A Note On Twistors}
%\label{twistorapp}

%The extension of the 2-spinor formalism to encompass conformal invariance is called \emph{twistor theory}, which is a vast area of research (see \emph{e.g.} \cite{penrose_rindler2}).  Twistor theory lies outside the scope of this paper, but for the sake of completeness, and comparison with other geometric algebra treatments of the subject, we will define them here.

%A \emph{valence-1 twistor} may be written in spinor form as (\emph{c.f.} \cite{2spinors})
%\bea
%\vec{\omega} = \vec{\omega}_0 - \pv{x} \tilde{\vec{\pi}}^\dagger_0 ,
%\eea
%where $\pv{x}$ is a position paravector with respect to an arbitrary origin, while $\vec{\omega}_0$ and $\vec{\pi}_0$ are constant two-component spinor fields \cite{penrose_rindler2}.  (Penrose and Rindler use the symbols $\mathring{\vec{\omega}}$ and $\mathring{\vec{\pi}}$, which we reject for typographic clarity.)  Algebraically is a general multivector in $\Cl^+_{1,3}$, and is similar in form to \eq{dirac2spin}, albeit with a specific dependence on position.  Twistors are designed with conformal invariance in mind, and as such are indeed related to Dirac spinors, whose inner product we already established to be conformally invariant.

%Although twistors are not studied as extensively now as in the past, some theorists still find them useful for calculating gauge boson amplitudes (reference).  In addition, they may play a role in supersymmetry and string theory \cite{witten03}.

\end{appendix}

\bibliography{spinors}

\end{document}